\documentclass[usenatbib]{mnras}
\usepackage{newtxtext,newtxmath}
\usepackage{pdflscape}
\usepackage[T1]{fontenc}
\usepackage{ae,aecompl}
\usepackage{placeins}
\usepackage{enumitem}

\usepackage{graphicx}	
\usepackage{amsmath}	
\usepackage{caption} 
\usepackage{tablefootnote}

\usepackage{soul}	
\setstcolor{red}
\setul{}{}		        

\usepackage{xcolor}  
\definecolor{ForestGreen}{RGB}{34,139,34}

\def\sun{\relax \ifmmode {\mbox H\,{\scshape ii}}\else H\,{\scshape ii}\fi}
\def\hii{\relax \ifmmode {\mbox H\,{\scshape ii}}\else H\,{\scshape ii}\fi}
\def\nii{\relax \ifmmode {\mbox N\,{\scshape ii}}\else N\,{\scshape ii}\fi}
\def\oii{\relax \ifmmode {\mbox O\,{\scshape ii}}\else O\,{\scshape ii}\fi}
\def\oiii{\relax \ifmmode {\mbox O\,{\scshape iii}}\else O\,{\scshape iii}\fi}
\def\sii{\relax \ifmmode {\mbox S\,{\scshape ii}}\else S\,{\scshape ii}\fi}
\def\ha{\relax \ifmmode {\mbox H}\alpha\else H$\alpha$\fi}
\def\hb{\relax \ifmmode {\mbox H}\beta\else H$\beta$\fi}

\def\deg{\hbox{$^{\circ}$}}         
\defcitealias{Strom2018}{ST18}    
\defcitealias{Strom2017}{ST17}    
\defcitealias{Amorin2010}{A10}
\defcitealias{Pilyugin2012}{P12}
\defcitealias{PM2021}{PM21}
\defcitealias{LA2020}{LA20}
\defcitealias{Hayden2021}{HP21}
\defcitealias{PMC09}{PM09}
\defcitealias{LS2015}{LS15}

\title[New calibrations for estimating the N/O ratio]{New calibrations for estimating the N/O ratio in \hii\ regions}

\author[E. Florido et al.]{
Estrella Florido,$^{1,2}$\thanks{E-mail: estrella@ugr.es}
Almudena Zurita,$^{1,2}$
and Enrique P\'erez-Montero$^{3}$
\\
$^{1}$Dpto. de F\'{\i}sica Te\'orica y del Cosmos, Campus de Fuentenueva, Edificio Mecenas, Universidad de Granada, E-18071 Granada, Spain\\
$^{2}$Instituto Carlos I de F\'{\i}sica Te\'orica y Computacional, Facultad de Ciencias, E-18071 Granada, Spain\\
$^{3}$Instituto de Astrof\'{\i}sica de Andaluc\'{\i}a, Camino Bajo de Hu\'etor s/n, Aptdo. 3004, E-18080 Granada, Spain
}

\date{Accepted XXX. Received YYY; in original form ZZZ}

\pubyear{2021}

\begin{document}
\label{firstpage}
\pagerange{\pageref{firstpage}--\pageref{lastpage}}
\maketitle

\begin{abstract}
  We use a sample of 536 \hii\ regions located in nearby spirals, with an homogeneous determination of their $T_e$-based abundances, to obtain new empirical calibrations of the   N2O2, N2S2, O3N2, and N2  strong-line indices to estimate the nitrogen-to-oxygen abundance ratio when auroral lines are not detected. All indices are strongly correlated with the $T_e$-based $\log$(N/O)  for our \hii\ region sample, even more strongly than with  $12+\log$(O/H). N2O2 is the most strongly correlated index, and the best fit to the $\log$(N/O)-N2O2 relation is obtained with a second-order polynomial. The derived relation has a low dispersion ({\em rms}$<$0.09~dex), being valid in the range $-1.74 < $ N2O2 $< 0.62$ (or   $-1.81 < $ $\log$(N/O) $< -0.13$). We have compared our calibration with previous ones and have discussed the differences between them in terms of the nature of the objects used as calibrators.

\end{abstract}



\section{Introduction}
Most of the chemical elements heavier than hydrogen and helium are formed by stars, which, at different stages of their evolution, expel part of them into the interstellar medium (ISM), enriching it with {\em metals}. The metal content in a galaxy changes with time depending on its overall star formation, but also on processes such as gas inflows of pristine gas or gas outflows and  stripping, that alter the metal concentration relative to hydrogen.
Therefore, the current chemical composition and the distribution of metals in the ISM of galaxies, provide crucial pieces of information to the key physical processes driving  galaxy evolution \citep[e.g.][]{Maiolino2019}. For this reason, oxygen, the most abundant metal in the Universe, has been extensively used as a tracer of the {\em metallicity} or 12+log(O/H) in the  gas-phase of galaxies \citep[e.g.][]{Searle1971,Pagel80,Bresolin2004,Kewley2010,Pilyugin2014,Bresolin2019}. Oxygen is a primary  element (its yield is independent of the initial chemical composition of the star), being produced by massive stars and ejected into the ISM in a short timescale via the explosion of type II SNe, a few Myr after the star was formed, thus increasing the O/H ratio in the ISM.

In addition to O/H, other abundance ratios such as  N/O are key tools to understand chemical evolution in galaxies.
However,  nitrogen production is much more complex than oxygen production, as it is produced in stars of all masses and it can have both primary and secondary origins \citep[e.g.][]{Vincenzo2016, Edmunds1978, Gavilan2006}.
 It is produced from C and O through the CNO cycle for the combustion of H in He, being most of it ($\sim74$\%) produced  in low- and intermediate-mass stars (LIM stars, $\lesssim8$M$_\odot$), according to chemical evolution models \cite[e.g.][]{Matteucci1986,Chiappini2003,Kobayashi2020}. The mean lifetimes of LIM stars are longer than those of more massive stars and, therefore, there is a time delay in the  enrichment of nitrogen in the ISM with respect to that of oxygen. The N/O abundance ratio of the ISM in a galaxy would then be sensitive to the
age of the galaxy, or to the time since most of its star formation has taken place, in the extreme case where all the star formation occurs in a single burst from the galaxy's gas (single stellar population) \cite[e.g.][]{Edmunds1978}. A more realistic scenario for spiral and irregular galaxies implies a continuous star formation, that makes the N/O abundance ratio very sensitive to their star formation history and efficiency \citep[][]{Molla2006,Vincenzo2016}.


The most commonly used metallicity indicators in the ISM, the O/H and N/O abundance ratios, are therefore complementary, with O/H being more fickle, more dependent on intermittent pollution by short-lived massive stars following star formation (hereinafter, SF) bursts, and modulated by the inflows and outflows of pristine  gas, while N/O is less sensitive to gas flows and more sensitive to the star formation history of the galaxy or the \hii\ region \citep[e.g.][]{Vincenzo2016,Molla2006,Koppen2005}. Indeed, an important diagnostic tool for constraining both chemical evolutionary processes and the origin of nitrogen is the O/H-N/O diagram, either for individual \hii\ regions \citep[e.g.][]{Alloin1979,Considere2000,Henry2000, Belfiore2017, Arellano2021}, as well as for all or part of the integrated light emitted by galaxies \citep[e.g.][]{AM2013, Amorin2010, Vincenzo2016, LA2020,Luo2020}.
In particular, our current improved observation capabilities, have permitted us to derive average O/H and N/O estimates  for high redshift galaxies, that compared with local estimates constrain the current theories about galactic evolution, or suggest new scenarios for that evolution \citep[e.g.][]{Maiolino2019, Vangioni2018, PM2021, PM2013, Sanders2018, Masters2014}.

The gas-phase N/O abundance ratio of galaxies has been found to correlate positively with the stellar mas of the galaxy \citep[e.g.][]{PMC09,PM2013}, as does the gas-phase O/H in the so-called mass-metallicity relation \citep[e.g.][]{Tremonti2004}. However, there are contradictory results regarding the dependence of the N/O-M$_\star$ relation on the star formation rate (SFR). According to \citet{PM2013}, it is independent of SFR, unlike O/H, but more recent results from \citet{Hayden2021} find the same anti-correlation of N/O with SFR as O/H. \citet{Hayden2021} explain the observed discrepancies, that can have important consequences for explaining the processes driving the fundamental metallicity relation as due to the choice of the empirical diagnostic used to derive N/O \citep[N2S2 vs. N2O2,][]{Hayden2021}.

Another advantage of N/O with respect to O/H is that the former has little dependence on the  gas electron temperature, $T_e$, as it is derived from a pair of collisionally excited lines ([\oii]$\lambda$3726,3729 and [\nii]$\lambda$6583) whose emissivities have similar dependence on $T_e$ \citep[e.g.][]{Skillman1998}, and it is also one of the most robust abundance relations against ionization correction factors to account for unobserved ionic species \citep[e.g.][]{Stasinska2002,AC2020,Esteban2018}.

Although it is not without problems \citep[e.g.][]{Stasinska2002,Bresolin2006}, the so-called {\em direct} or $T_e$-based method\footnote{Throughout this paper we will refer to the abundances obtained by this method indistinctly as {\em direct} or $T_e$-based abundances.} \citep[e.g.][]{Bresolin2004,Berg2020,PM2014} is considered the most accurate for determining the O/H and N/O abundance ratios in \hii\ regions or in the gas-phase of galaxies. The drawback is that it requires the detection of very faint collisionally excited emission lines, known as {\em auroral} lines, which allow the estimation of $T_e$, a necessary step to determine the corresponding ionic abundance. The detection of the auroral lines is more complicated at high metallicities, as an increase in metallicity has the effect of increasing the cooling in the nebula, and the relevant emission-line ratios remain unreachable even for 10m class telescopes \citep{Bresolin2006}. An alternative method is that based on measurements of optical metal recombination lines, where the line emissivities are only moderately dependent on $T_e$, but these lines are very faint and only detectable in the Milky Way and nearby galaxies \cite[e.g.][]{Esteban2020,Toribio2016,Garcia-Rojas2006}.

 These observational limitations have motivated the use of {\em strong-line} methods, first introduced by \citet{Pagel1979}, as an alternative to estimate the gas-phase oxygen abundance  when auroral lines and/or  optical metal recombination lines can not be detected. These methods are based on the calibration of ratios of strong and easily detectable emission lines of the nebula. The strong-line ratios are typically calibrated either using $T_e$-based metallicities \citep[e.g.][]{PP2004,B07, Marino13,  PG2016} or  metallicities predicted with photoionization models  \citep[e.g.][]{MG91,Kewley2002, Dopita2016}. Both photoionization models and methods based on metal recombination lines tend to estimate larger oxygen abundances than the $T_e$ method \citep[e.g.][]{Bresolin2009,AZ2021a,ETG2018} but see also \citet{Dors2011} and \citet{PM2014}.
  
 Strong-line methods for the derivation of O/H (metallicity) are extremely popular, and a high number of different calibrations and methods are available in the  literature \citep[see e.g.][for a compilation]{Maiolino2019}. Although these methods are less precise than the $T_e$-based method, they have been extremely useful for making metallicity estimates for large sets of \hii\ regions or  star-forming sites \cite[e.g.][]{SanchezMenguiano2018,AZ2021a} or for distant galaxies, especially at high redshift, where usually only a few emission lines are detected \cite[e.g.][]{Dopita2016,Brown2016}, leading to important results and scaling relations \cite[e.g.][]{Tremonti2004}.

Despite its relevance, the number of available strong-line methods for the derivation of N/O are much scarcer than those for O/H, and the ones available have not been as carefully tested as those to derive O/H. In the derivation of empirical strong-line methods, there are some minimum requirements, important for a reliable  calibration which are related to the properties of the calibration sample: (1) It must comprise a large number of objects with $T_e$-based abundances, (2) the methodology to derive abundances for all the targets must be homogeneous,  and (3) the calibration targets must cover a wide range of abundances and/or strong-line index values.

The aims of this work are to analyse some of the most frequently used strong-line methods to estimate the N/O abundance ratio in \hii\ regions and in the gas-phase of star-forming galaxies in the literature, and to derive new empirical strong-line methods to estimate N/O. We will make use of the recent compilation of emission-line fluxes for \hii\ regions in nearby galaxies done by \citet{AZ2021a}, for investigating the effect of galactic bars on the gas-phase abundance gradients of spirals \citep{AZ2021b}.

This paper is organised as follows. In Section~\ref{sample} we summaryse the \hii\ region sample properties. In Section~\ref{calib} we first analyse the strong-line ratios or indices that are suitable candidates to be indicators of the N/O abundance ratio in \hii\ regions, and then explore  their dependence with second order parameters. The comparison of our new empirical calibrations to derive N/O with previous methods is presented in Section~\ref{comparison}. We discuss the results in Section~\ref{discussion} and, finally, present our conclusions  in  Section~\ref{conclusions}.


\section{H{\sc II} region sample}
\label{sample}
This work is based  on a large sample of \hii\ regions for which emission-line fluxes were available in the literature and compiled by \citet{AZ2021a}. The sample comprises  2831 independent measurements of \hii\ regions\footnote{Some of these measurements correspond to the same \hii\ region, but observed by different authors. All different observations of the same target were kept as independent observations.} from 51 nearby ($<64$~Mpc) spiral galaxies with absolute B-band magnitudes between -22 and -17, and inclinations lower than 70$\deg$.

The compilation comprises celestial coordinates and emission-line fluxes, including those from auroral lines, when available in the original papers. The latter permitted the derivation of O/H and N/O abundance ratios with the $T_e$-based or {\em direct} method for 610 and 536 \hii\ regions, respectively, with a homogeneous  methodology\footnote{The collected emission line fluxes are not homogeneous, as they come from previous publications by different authors. However, we concentrated our collection on resolved spectroscopic data of \hii\ regions, and we estimated that the differences in the emission-line fluxes for a given region, as measured by different authors, are not larger than the typical observational error associated of emission-line flux measurements \citep[see section 3.1 in][]{AZ2021a}.}. The {\em direct} abundances range from 7.42 to 9.07 for 12+$\log$(O/H), and from -1.81 to -0.13 for $\log$(N/O).

We refer the reader to \citet{AZ2021a} for details on the galaxy and \hii\ region sample, and on the methodology to derive the chemical abundances from the compiled emission-line fluxes.

In this paper we take advantage of this compilation with a double purpose: (1) to analyse some of the most frequently used strong-line methods to estimate the N/O abundance ratio in \hii\ regions and in the gas-phase of star-forming galaxies, and (2) to derive new empirical strong-line methods for the derivation of N/O in \hii\ regions. 
Our calibrating sample comprises the 536 \hii\ regions for which the $T_e$-based N/O abundance ratio was derived in \citet{AZ2021a}. This calibrating sample has the advantage of uniformity, as it is composed by \hii\ regions within spiral galaxies alone, in contrast with previous samples that included \hii\ regions from different galaxy types and/or integrated fluxes from galaxies \citep[e.g.][]{PMC09,Pilyugin2012,Li2006}. 


\section{Strong-line ratios to derive N/O}
\label{calib}

The difficulties for detecting reliably the faint auroral lines needed for the determination of $T_e$, motivated the development of alternative methods to estimate chemical abundances from strong nebular emission lines. The most widely used methods involve fluxes from the [\nii]$\lambda$6583, [\oiii]$\lambda$5007, [\oii]$\lambda$3727 and/or [\sii]$\lambda\lambda$6717,6731 emission lines, besides the \ha\ and \hb\ Balmer series lines, as these are easily detected in the optical range in spectroscopic observations of nearby galaxies with medium-size telescopes from Earth. The number of different strong-line ratios and calibrations for the derivation of O/H is enormous, while this number is much limited for the derivation of N/O \citep[see e.g.][for a review]{Maiolino2019}. One of the drawbacks of some of the strong-line methods in use for deriving O/H, is their dependence on other parameters (density, ionization parameter,...). In particular, those including the [\nii]$\lambda$6583 emission line introduce a dependence of the O/H derived abundances with the N/H or N/O abundance ratio \citep[e.g.][]{PM2005,PMC09,AZ2021a}. 

We have selected some of the most widely used metallicity strong-line indices that include in their definition the [\nii]$\lambda$6583 emission-line flux:

\begin{itemize}
	\item N2 = $\log$([\nii]$\lambda$6583/\ha)
	\item O3N2 = $\log$(([\oiii]$\lambda$5007/\hb)/([\nii]$\lambda$6583/\ha))
	\item N2O2 = $\log$([\nii]$\lambda$6583/[\oii]$\lambda$3727)
	\item N2S2 = $\log$([\nii]$\lambda$6583/[\sii]$\lambda$6717,6731) 
\end{itemize}

The N2, O3N2 and N2O2 indices are widely used for the derivation of O/H \citep[see e.g.][respectively]{Shapley2005,SanchezMenguiano2018,Ho2015}. The N2 index \citep{Storchi-Bergmann1994,Denicolo2002} has the advantage of using very close-by lines in the spectra, which minimizes the effect of differential dust-extinction and requires a very small spectral coverage in observations, but it has the disadvantages that it saturates at high metallicity \citep{Baldwin1981}, its behaviour is not linear at low metallicities \citep{ML2014} and it depends strongly on the ionization parameter \citep[e.g.][]{PM2005}.  O3N2 was first proposed by \citet{Alloin1979}, and it is a very popular method for deriving  oxygen abundances, since it is also useful in surveys with limited spectral coverage, although it is strongly dependent on the \hii\ region ionization parameter \citep[e.g.][]{Ho2015,AZ2021a} and it saturates at low oxygen abundances values \citep[12+$\log$(O/H)$\lesssim8$,][]{PMC09}.

The N2S2 index also makes use of close-by in wavelength emission lines, but it considers the [\sii]$\lambda\lambda$6717,6731 doublet, which is fainter and therefore more difficult to detect.  It was first proposed as a metallicity indicator for \hii\ regions by \citet{Viironen2007}, and later on by \citet{Dopita2016} for high-redshift galaxies. \citet{PMC09} proposed it as an indicator of the N/O ratio on ionized nebulae.
The N2O2 index, suggested by \citet{Kewley2002},  is the one most affected  by differential extinction and by flux calibration uncertainties, but it is a frequently used metallicity indicator for \hii\ regions with $12+\log$(O/H)$\gtrsim7.6$ \citep{Dopita2000}, and it is also a useful tool to derive the N/O abundance ratio \citep{PMC09}.  Both N2O2 and N2S2 are virtually independent on the ionization parameter  \citep[e.g.][]{Dopita2000,Nagao2006,AZ2021a}.

The strengths, weaknesses and caveats of the use of these indices for the derivation of metallicities have been widely discussed in the aforementioned references and/or evaluated against the more reliable $T_e$-based method,  and to a much lesser degree for the derivation of N/O \citep[see also][]{PM2005,KE2008,LE2010,Maiolino2019,Arellano2020,AZ2021a}.

We have used our sample of \hii\ regions (Section~\ref{sample}) to analyse the relation between the N2, O3N2, N2O2 and N2S2 indices and the  $T_e$-based N/O abundance ratio. As a first step, we derive the Spearman's rank  correlation coefficient  ($\rho$)  between the \hii\ region values for each index, and the  corresponding $T_e$-based $\log$(N/O) abundance ratio. The results are shown  in  Table~\ref{tab:linearfit}, together with the number of data values ($N$) involved in each relation. All indices show a strong correlation with $\log$(N/O), with values of $\rho$ greater than 0.78, but the strongest correlation is found for the N2O2 index, with $\rho$ = 0.95. For comparison, we show in the same table the correlation coefficient of the same indices with the $T_e$-based oxygen abundance, for which they are more widely used. The Spearman's rank  correlation coefficient for $12+\log$(O/H) is lower than for  $\log$(N/O), indicating moderate correlations, with $\rho$ values around 0.6 for all indices. 

\begin{table}
\begin{minipage}{0.5\textwidth}
	\centering
	\caption{Spearman rank  correlation coefficient  ($\rho$)  for both the $T_e$-based $\log$(N/O) and $12+\log$(O/H) abundances as a function of the four selected indices. $N$ is the number of \hii\ regions used in each inspected relation.}
	\label{tab:linearfit}
	\begin{tabular}{l|rrr|rrr} 
	  \hline
             & \multicolumn{3}{|l|}{$\log$(N/O)}  & \multicolumn{3}{|r}{$12+\log$(O/H)}  \\
           \hline
		Index & $N$  & $\rho$ & & & $N$ & $\rho$ \\
                \hline 
                 N2   & 536  & 0.88  & & & 544  &  0.55 \\
                 O3N2 & 530  & -0.85 & & & 538  & -0.56 \\
                 N2O2 & 536  & 0.95  & & &  544  & 0.62\\
                 N2S2 & 530  & 0.78  & & & 538  & 0.58 \\		
	         \hline 
	\end{tabular}
\end{minipage}
\end{table}

In the next subsections we further explore the use of the four strong-line indices as indicators of $\log$(N/O) for our \hii\ region sample, with more emphasis on the N2O2 index that shows the strongest correlation with the  $T_e$-based nitrogen-over-oxygen abundance ratio.

\subsection{N2O2}
\begin{figure}
 \centering
	\includegraphics[width=0.47\textwidth]{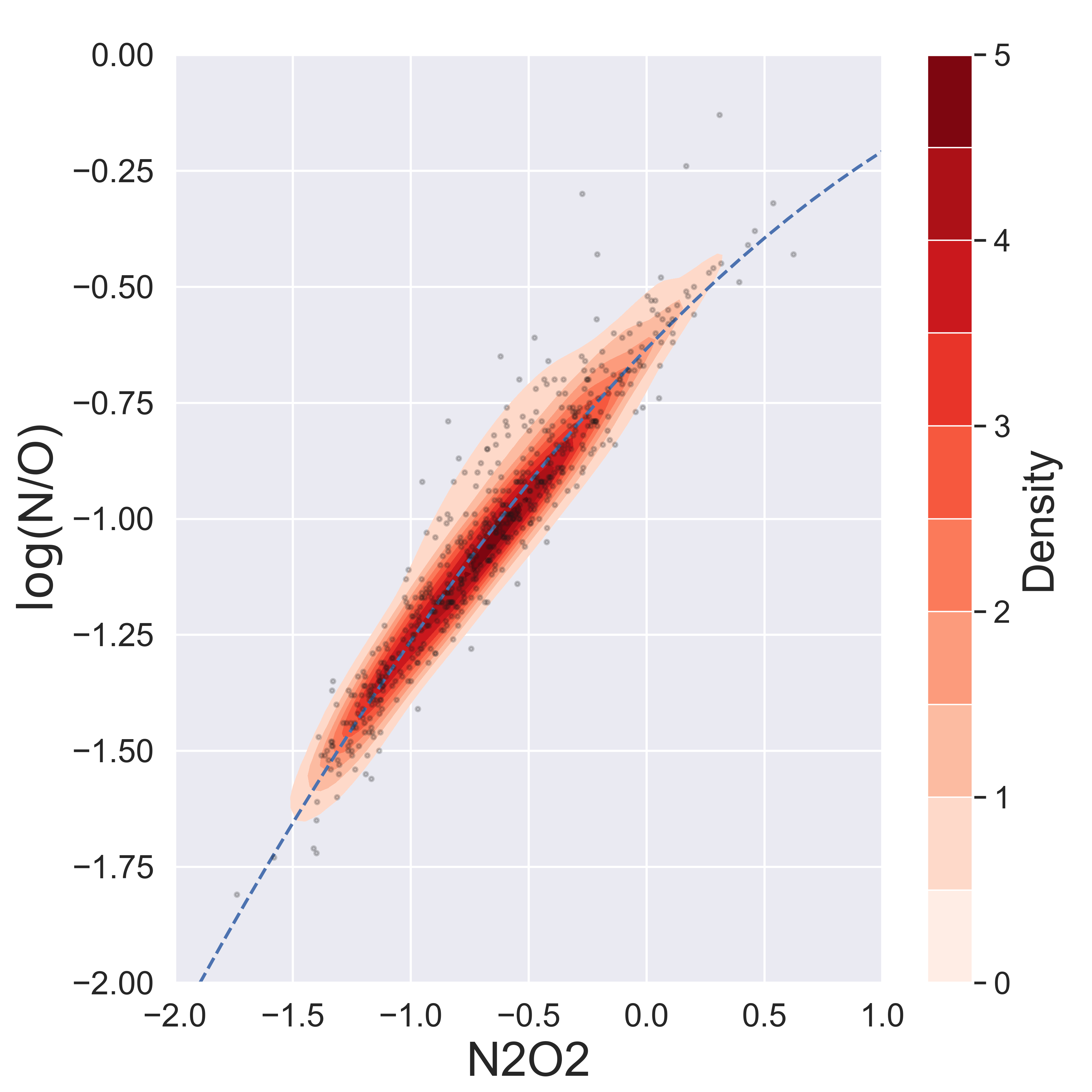}
	\caption{$\log$(N/O) for \hii\ regions as derived from the {\em direct} method \citep{AZ2021a} vs. the N2O2 index. Both individual data points (black dots) and a density map from the same data set are shown. The dashed-blue line shows the best fit to the data points (see text for details). Typical uncertainties in $\log$(N/O), not plotted for clarity, are $\sim$ 0.17 dex}
	\label{fig-ajustesNO-N2O2}
\end{figure}

Fig.~\ref{fig-ajustesNO-N2O2} shows the $\log$(N/O) abundance ratio derived from  the $T_e$-method against the N2O2 index. Individual \hii\ region values are indicated with black dots. An underlying  density map for the same data set is shown in red colors, that allows a better inspection of the distribution in areas of the plot with a high overlapping of data-points (darker red).

We have derived the function that best fits the  $T_e$-based $\log$(N/O) - N2O2 relation. Several fitting techniques have been explored that include the use of  bayesian statistical methods, weighted/unweighted fits considering uncertainties in both parameters, the use of different degrees in the polynomial fitting and/or the derivation of the best-fitting function for average values within N2O2 and $\log$(N/O) bins, rather than for the individual data points.  For a first evaluation of the goodness of the fit we have calculated the median value of the fitting residuals and their standard deviation.

\begin{figure*}
\centering
\includegraphics[width=1.\textwidth]{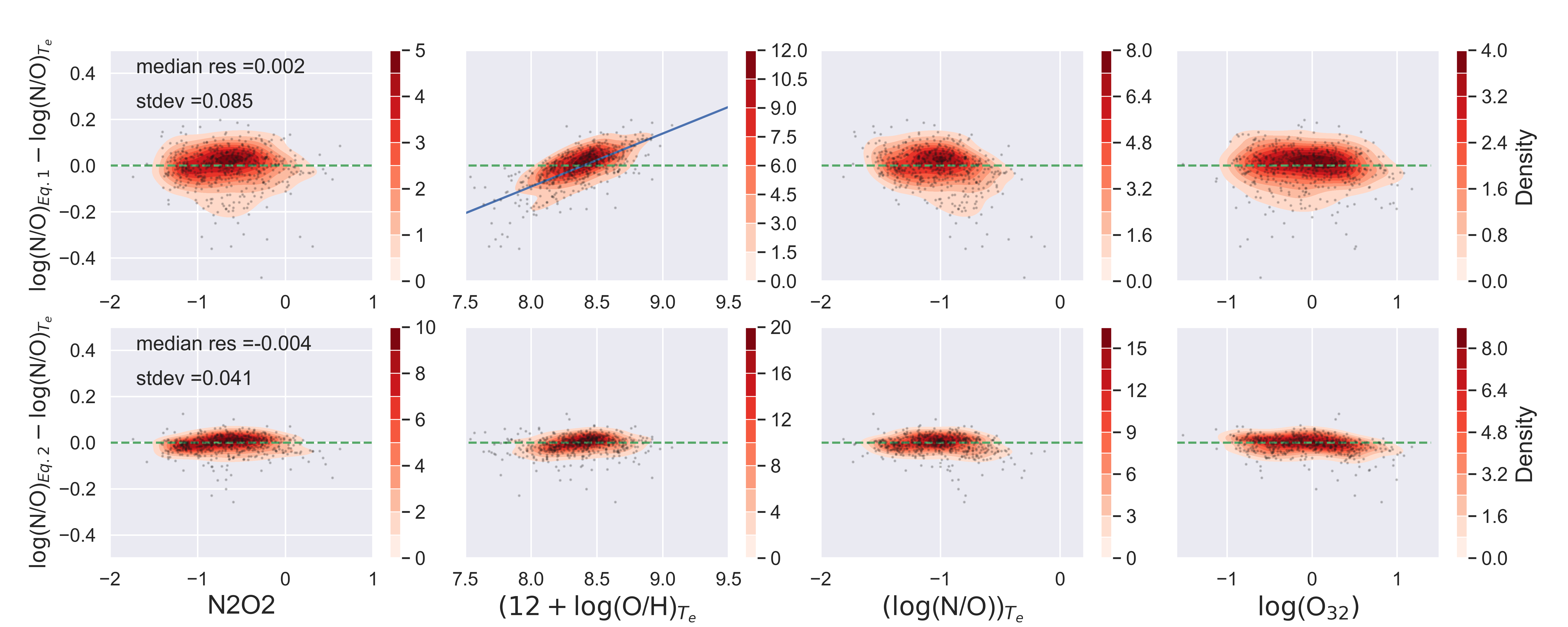}
       \label{b}    
 \caption{Difference between the N/O abundances derived from our best fittings and the corresponding $T_e$-based value (residuals) as a function of the N2O2 index, the $T_e$-based O/H and N/O abundances, and $\log$~O$_{32}$, for the best fitting in Eq.~\ref{fit1} and Eq.~\ref{fit2} in the top and bottom panels, respectively. An underlying density map for the same data set is shown in red colors in addition to the individual data points (black dots). The median value of the residuals and their standard deviation for each fitting is indicated in the corresponding left hand side panel. The green-dashed straight line mark the zero residuals line. The blue straight line shows the linear correlation between the residuals from Eq.~\ref{fit1} and the $T_e$-based $12+\log$(O/H).}
\label{fig-ajustesNO-N2O2-res}
\end{figure*}

The best-fitting to the data  is obtained  with a second-order polynomial, with weights equal to the reciprocal of the squared uncertainties in  $\log$(N/O)  and yields 
\begin{equation}
\label{fit1}
 \begin{array}{rl}
     \log\mbox{(N/O)}_{N2O2} = & (-0.102 \pm 0.018) \times \rm{N2O2}^2    \\
                        & +(0.528 \pm 0.019) \times \rm{N2O2} - (0.634 \pm 0.006) \\
 \end{array}
\end{equation}
It is shown with a blue dashed-line in Fig.~\ref{fig-ajustesNO-N2O2}. The median value of the residuals is 0.002$\pm$0.004~dex and their standard deviation is 0.085~dex. 
A careful analysis of the fitting residuals is shown in the top panels of Fig.~\ref{fig-ajustesNO-N2O2-res} where these are plotted as a function of the N2O2 index, the $T_e$-based metallicity,  the $T_e$-based $\log$(N/O) and $\log$~O$_{32}$ ($=\log$([\oiii]$\lambda\lambda4959,5007$/[\oii]$\lambda3727$)) as a proxy for the ionization parameter in the \hii\ regions. It can be clearly seen that the residuals from our fit (Eq.~\ref{fit1}) show a positive correlation with the $T_e$-based oxygen abundance, whereas no correlation is observed with either the  $T_e$-based $\log$(N/O) or $\log$O$_{32}$.

In order to correct for this, we have tested a new fitting function taking into account both the N2O2 index and the $T_e$-based O/H abundance as independent variables, and using the reciprocal of the squared uncertainties in $\log$(N/O) as weights. The resulting function is:

\begin{equation}
\label{fit2}
\begin{array}{rl}
     \log\rm{(N/O)}_{cor} = & (-0.160 \pm 0.011) \times \rm{N2O2}^2    \\
                        & +(0.59 \pm 0.02) \times \rm{N2O2} + (2.20 \pm 0.08) \\
                        & - (0.330 \pm 0.009) \times  (12+\log\rm{(O/H)}) \\
\end{array}
\end{equation}
  The fitting residuals from Eq.~\ref{fit2} are shown in the  bottom panels of Fig.~\ref{fig-ajustesNO-N2O2-res}.  There are no significant dependences of the residuals with the oxygen abundance or the ionization parameter  of the \hii\ regions. The dispersion of the  residuals has been reduced with the fitting of Eq.~\ref{fit2} from 0.085 to 0.041~dex. Their median value is however slightly  higher now than with the  fitting from Eq.~\ref{fit1}, $-0.004\pm0.002$~dex, but  both values are comparable if we consider error bars, and are, in any case, smaller than the data dispersion. 

\begin{figure}
\centering
\includegraphics[width=0.47\textwidth]{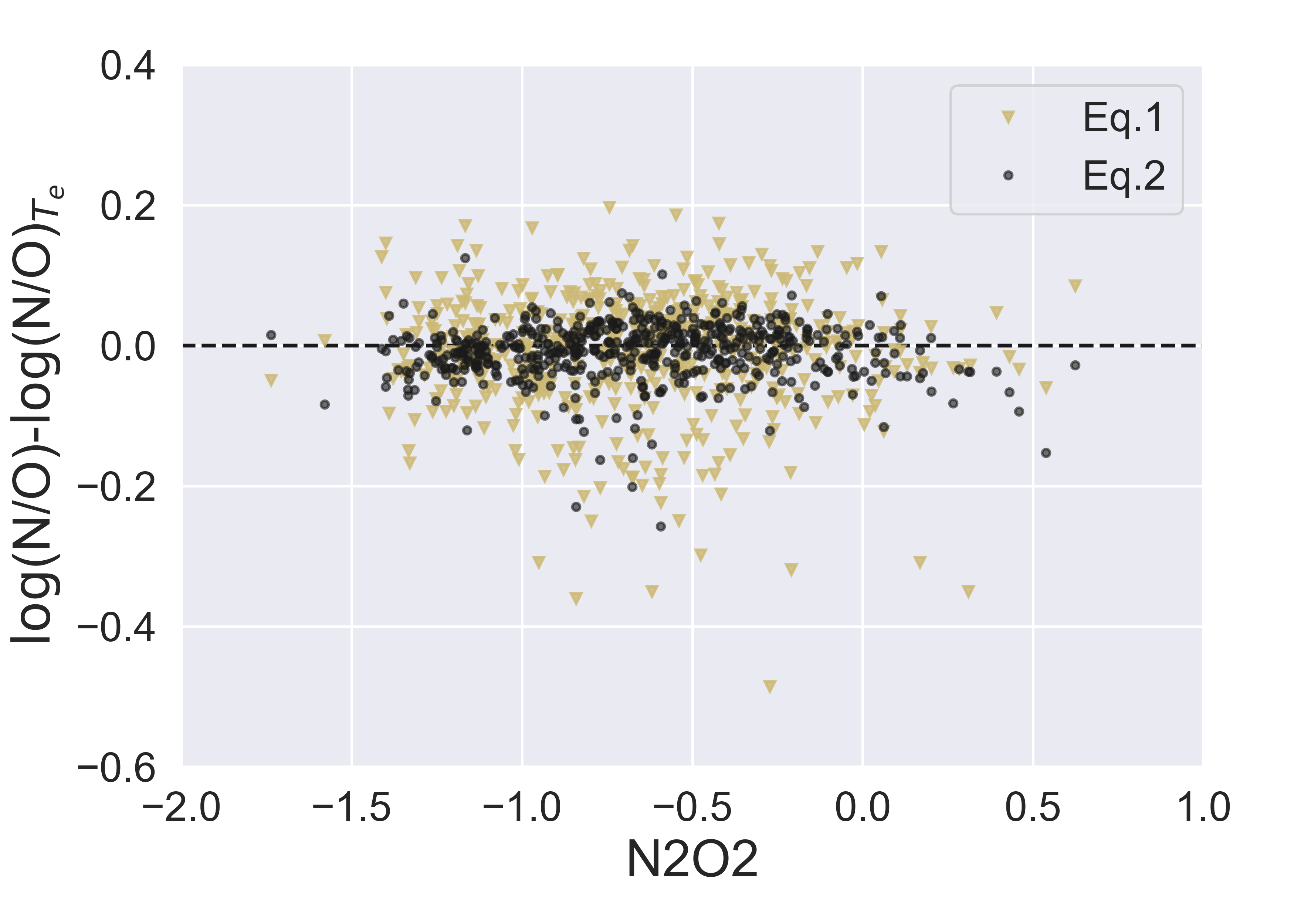}
\caption{Fitting residuals for the two empirical calibrations of the N2O2 index for the derivation of  $\log$(N/O), as a function of  N2O2 for our \hii\ region calibrating sample. Yellow triangles and black dots correspond to residuals from the fitting functions in Eq.~\ref{fit1} and Eq.~\ref{fit2}, respectively.}
\label{fig-NO-N2O2-inicial-final}
\end{figure}

The improvement reached with this new fitting in terms of the reduction of residuals dispersion and its dependence on the $T_e$-based $12+\log$(O/H)
is better seen in Fig.~\ref{fig-NO-N2O2-inicial-final},  where we plot the residuals from the two best-fittings (Eq.~\ref{fit1} and Eq.~\ref{fit2}) as a function of the N2O2 parameter with yellow triangles and black dots,  respectively. It is important to note that although Eq.~\ref{fit2} yields a better fitting to the data, the empirical calibration of N2O2 to derive  $\log$(N/O) given in Eq.~\ref{fit1} is already very good. Eq.~\ref{fit2} requires knowledge of the $T_e$-based 12+$\log$(O/H). Its usefulness is then limited to cases where the electron temperature and the $direct$ oxygen abundance can be determined, but the N/O abundance ratio cannot be derived, presumably in limited spectral range observations, which do not include the [\nii]$\lambda$6583 emission line. However, even for these objects, Eq.~\ref{fit1} would already yield a rather acceptable estimation of log(N/O).
In fact,  the mean (0.0002 dex) and  median (-0.008 dex)  difference in the  $\log$(N/O) values  derived from Eq.~\ref{fit1} and  Eq.~\ref{fit2}  are much smaller than the  typical  median  values (0.17 dex) of the uncertainties in the  $T_e$-based $\log$(N/O) abundances for the \hii\ regions in our sample. 

The range of validity of the two fittings, the  number of \hii\ regions used for these empirical calibrations of N2O2, and the median value and the standard deviation of the fitting residuals are summarized in the first two lines of  Table~\ref{tab:ajustes}.

\subsection{N2S2, O3N2 and N2}
\begin{figure*}
\centering
      \includegraphics[width=1.\textwidth]{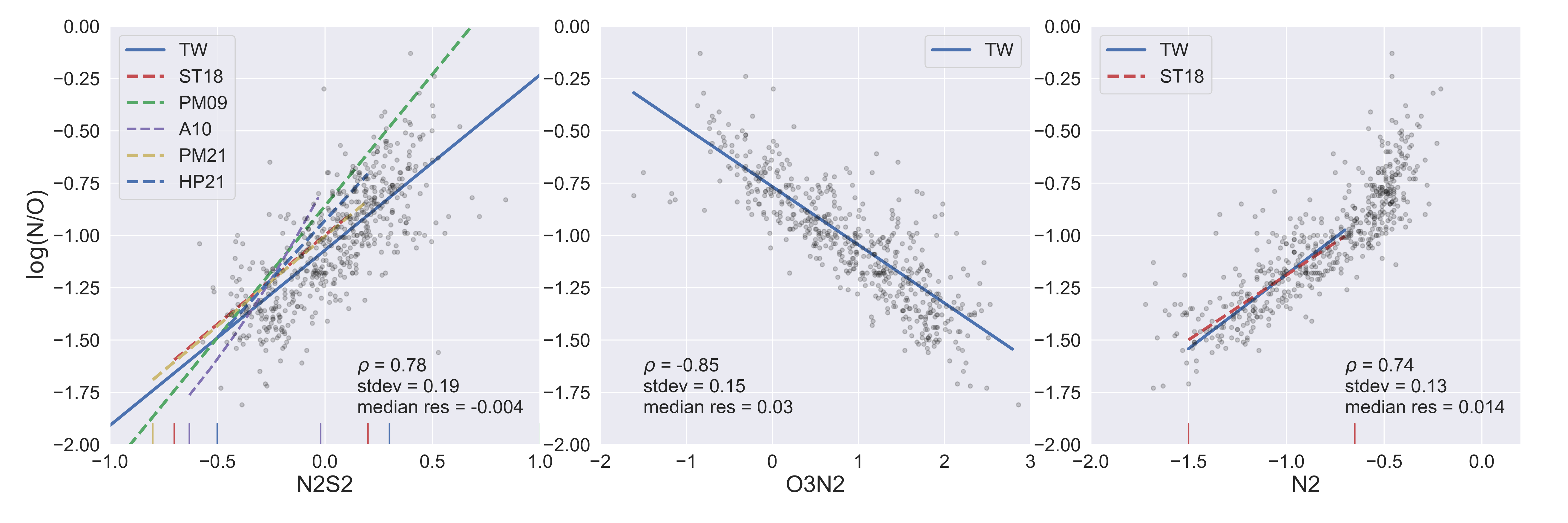}  
   	\caption{$T_e$-based $\log$(N/O) abundance ratio vs. the strong-line ratios N2S2, O3N2 and N2. The blue straight line shows our best linear fitting to the data (TW).  The Spearman's rank correlation coefficient  is shown in the bottom part of the plots, together with the standard deviation and median value of the fitting residuals. Calibrations of these indices obtained by ~\citet{Strom2018},  ~\citet{PMC09}, ~\citet{Amorin2010}, ~\citet{PM2021} and \citet{Hayden2021} are shown in the corresponding panels with green, red, purple, yellow and blue dash lines, respectively.
  Vertical segments  in the bottom part of the panels indicate the   validity range limits  for each calibration, with the corresponding color. The blue straight line in the right-hand panel shows the best fitting  (TW) obtained for the N2 index for values of N2 between -1.5 and -0.65. See text for details.}
	\label{NO-N2-N2S2-O3N2}
\end{figure*}

The N2S2, O3N2 and N2 indices are widely used for the derivation of metallicities. However, although these indices are less strongly correlated with $\log$(N/O) than  N2O2  (Table~\ref{tab:linearfit}), their correlation with N/O is also  strong  ($\rho\sim 0.78 - 0.88$) and, in any case, much stronger than their correlation with $12+\log$(O/H),  with  $\rho\sim 0.55 - 0.58$.  We explore here  the usefulness of N2S2, O3N2 and N2 as empirical indicators of N/O in our \hii\ region sample. Fig.~\ref{NO-N2-N2S2-O3N2} shows  $T_e$-based $\log$(N/O) abundances from \citet{AZ2021a} as a function of N2S2, O3N2 and N2.
The figure shows, as a first approach, the best linear fittings (blue straight lines). The dispersion of the fitting residuals is slightly worse for N2S2 ($\sim0.19$ dex) than for O3N2 and N2 ($\sim0.14$  dex), but the median value for the residuals is very close to cero (-0.004 dex).

The O3N2 and N2 indices are known to be strongly dependent  on the  \hii\ region ionization parameter ~\citep[e.g.][and references therein]{LS2011,Marino13,PMC09,PM2005}.
 In fact, we have seen this dependence for the residuals of the best-fittings shown in Fig.~\ref{NO-N2-N2S2-O3N2}. Therefore, we have performed a second fitting of $T_e$-based $\log$(N/O) values for the \hii\ regions of our sample as a function of  two independent variables:  the index (either O3N2 or N2) and $\log$(O$_{32}$), the later as a proxy for the ionization parameter.  The resulting fits reduce considerably the dispersion of the residuals down to $\sim0.09$~dex for the two indices, while the median value for the residuals remains similar for O3N2 as for the previous fitting, and is reduced down to -0.003~dex for N2. As N2 saturates at high abundances (N2$\gtrsim$-0.6), in Fig.~\ref{NO-N2-N2S2-O3N2} we have restricted the fitting to \hii\ regions with N2 values in the range  $-1.5 < N2 < -0.65$ (294 out of 536 regions).

Table~\ref{tab:ajustes} shows the best-fitting parameters for the $\log$(N/O) in our \hii\ region sample as a function of the different strong-line ratios (N2O2, N2S2, O3N2 and N2), including those for the fits considering second order dependences on  $12+\log$(O/H), for N2O2, and on  $\log$(O$_{32}$) for O3N2 and N2. The range of validity of the different empirical calibrations of these indices for the derivation of $\log$(N/O) is also given.

In summary, the four indices show a strong correlation with $\log$(N/O). Our best fittings to the $T_e$-based \hii\ region N/O abundances indicate that N2O2 is the best tracer of N/O, given the small dispersion of the fitting residuals and the relatively wide range of validity for this empirical calibration. O3N2 and N2 also show a small dispersion when their dependence on the ionization parameter is taken into account, but their range of validity is smaller, especially for N2.

\begin{table*}
\begin{minipage}{\textwidth}
\centering
\caption{Best-fitting parameters to the $T_e$-based $\log$(N/O) as a function of the  strong-line indices N2O2, N2S2, O3N2 and N2, for the empirical calibrations derived in this work. Columns (1) and (2)  show the independent variables for the fitting; (3) resulting empirical calibration; (4) validity range for the calibration given in column  (3); (5) Spearman's rank correlation coefficient between $\log$(N/O) and the strong-line index in column (1), $x_1$, in this index range; (6) standard deviation and (7) median value of the fitting residuals, and (8) shows the number of \hii\ regions used in each fitting. Note that the number of \hii\ regions used for the calibration of the N2 index is much smaller than for the other indices, due to its reduced range of validity.\label{tab:ajustes}}
\begin{tabular}{lclcrccc} 
\hline
$x_1$ & $x_2$ & $\log$(N/O)=       & validity range  & $\rho$ & Residuals  & Median   & N\\
     &       &                    &  (dex)           &        &   stdev   & residuals &  \\

     &       &                    &   &        &   (dex) & (dex) &  \\
(1)  &  (2)  & (3)                &  (4)   &  (5)   &   (6) & (7) &(8) \\
\hline
\hline
N2O2 &   -   &   $(-0.102\pm0.018){x_{1}}^2 + (0.528 \pm 0.019) x_1$                   & $-1.74 <x_1 < 0.62$  & 0.95 & 0.085 & 0.002 & 536\\
&           &      $- (0.634 \pm 0.006)$                                               &                      &     &       &   &     \\ \hline
N2O2 & 12+$\log$(O/H) &   $(-0.16 \pm 0.01)x_1^2 +(0.59 \pm 0.02) x_1$               &  $-1.74 < x_1 < 0.62$ &0.95 & 0.041 & -0.004 & 536\\
     &              &   $+(2.20 \pm 0.08) -  (0.330 \pm 0.009) x_2$                  &                      &     &       &  &  \\\hline                       
N2S2 &     -           &   $(0.84 \pm 0.03)  x_1  - (1.071 \pm 0.008)$                  &  $-0.58 < x_1 < 0.84$ &0.78 & 0.19 & -0.004 & 530\\ \hline
O3N2 &     -           &   $(-0.28 \pm 0.01)  x_1  - (0.77 \pm 0.01)$                   &  $-1.61 < x_1 < 2.86$ &-0.85 & 0.15 & 0.03 & 530\\\hline
O3N2 &$\log$(O$_{32}$)   &   $(-0.73 \pm 0.01)  x_1  - (0.29 \pm 0.02) + (0.74 \pm 0.02)  x_2$ & $-1.61 < x_1 < 2.86$ &-0.85 & 0.09 & 0.025 & 530\\ \hline
N2   &          -      &   $(0.71 \pm 0.04)  x_1  - (0.48 \pm 0.04)$                         & $-1.5 < x_1 < -0.65$ &0.74 & 0.13 & 0.014 & 294\\  \hline
N2   &$\log$(O$_{32}$)   &   $(0.85 \pm 0.03)  x_1  - (0.40 \pm 0.03) + (0.33 \pm 0.02)  x_2$  & $-1.5 < x_1 < -0.65$ &0.74 & 0.09 & -0.003 & 294\\\hline
\end{tabular}
\end{minipage}
\end{table*}


\section{Comparison with previous strong-line calibrations for N/O estimates}
\label{comparison}
We have made a rather exhaustive, although possibly not complete, compilation from the literature of the most recent and frequently used methods to derive the gas-phase N/O abundance ratio. We have paid especial attention to the type of calibrating objects and to the ranges of validity for each analysed method. All of them, in addition to our derived calibrations, appear in   Table~\ref{tab:otros}, where we also show the most relevant pieces of information for each of these methods.  

In the following subsections we will carefully compare all the collected methods with our derived strong-line calibrations (Section~\ref{calib}) to estimate $\log$(N/O). Our comparison will be based on two elements: (1) the derived empirical parametrizations, and (2) the standard deviation and the median value of the difference between the $T_e$-based $\log$(N/O) values and those obtained from each strong-line method for our \hii\ region sample (columns 7 and 8 in Table~\ref{tab:otros}).

\subsection{Methods based on the N2O2 index}
\label{methods-N2O2}
\begin{figure}
 \centering
	\includegraphics[width=0.5\textwidth]{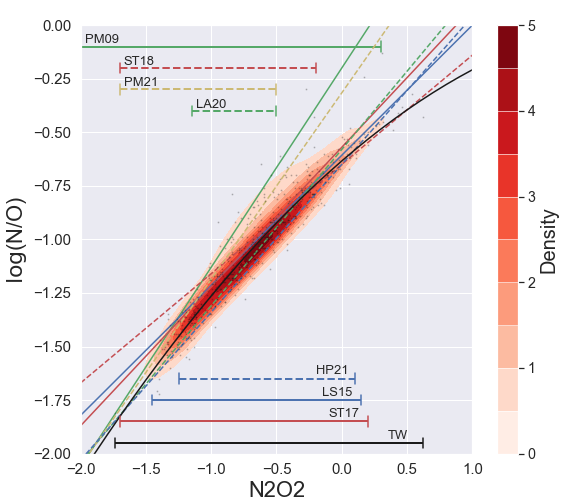}
	\caption{Direct-method N/O abundance vs. strong-line ratio N2O2: dots are individual values and in red is represented the density map. Fit from Eq. 1 is shown in black continuous line (TW). Besides, we show other calibrations of N2O2 compiled from the literature: ~\citet{PMC09} (PM09), ~\citet{LS2015} (LS15), ~\citet{Strom2017} (ST17), \citet{Strom2018} (ST18), \citet{LA2020} (LA20), \citet{PM2021} (PM21) and \citet{Hayden2021} (HP21).}
	\label{fig-ajustesNO-otros}
\end{figure}

N2O2 is the most commonly used index for estimating the ionized gas-phase N/O abundance ratio. Some of the most frequently used calibrations of this index derived by previous work are summarized in the top rows of Table~\ref{tab:otros}. Their corresponding analytical expressions have been used to overplot them in Fig.~\ref{fig-ajustesNO-otros}, on the $\log$(N/O) {\it versus} N2O2 diagram for our \hii\ region sample.

Except for the calibration performed by \citet{Strom2018} and \citet{LS2015}, hereinafter  \citetalias{Strom2018} and \citetalias{LS2015}, respectively,
all the rest have been obtained empirically from determinations of the N/O abundance ratio from the $T_e$-based method. However, the nature of the targets used as calibrators changes from one calibration to another. The analytical calibrations by \citetalias{LS2015} and \citet{Strom2017}, hereinafter \citetalias{Strom2017}, are the only ones based exclusively on \hii\ regions. The calibration by \citetalias{LS2015} is obtained from N/O abundances derived from an empirical determination of the $T_e$ for 48 \hii\ regions of the galaxy pair NGC1510/NGC1512, with N2O2 values in the range -1.45 $-$ 0.15 dex, very similar to the range covered by our data sample. Their empirical calibration departs ($\lesssim 0.18$ dex) from ours only for low values of N2O2 ($\le -1.3$ dex). The median difference between $T_e$-based $\log$(N/O) and the $\log$(N/O) values derived from the  \citetalias{LS2015} calibration for our \hii\ region sample is 0.04 dex, with a standard deviation of 0.09 dex, very similar to the dispersion of the residuals for our calibration (Eq.~\ref{fit1}).

\begin{figure}
 \centering
	\includegraphics[width=0.47\textwidth]{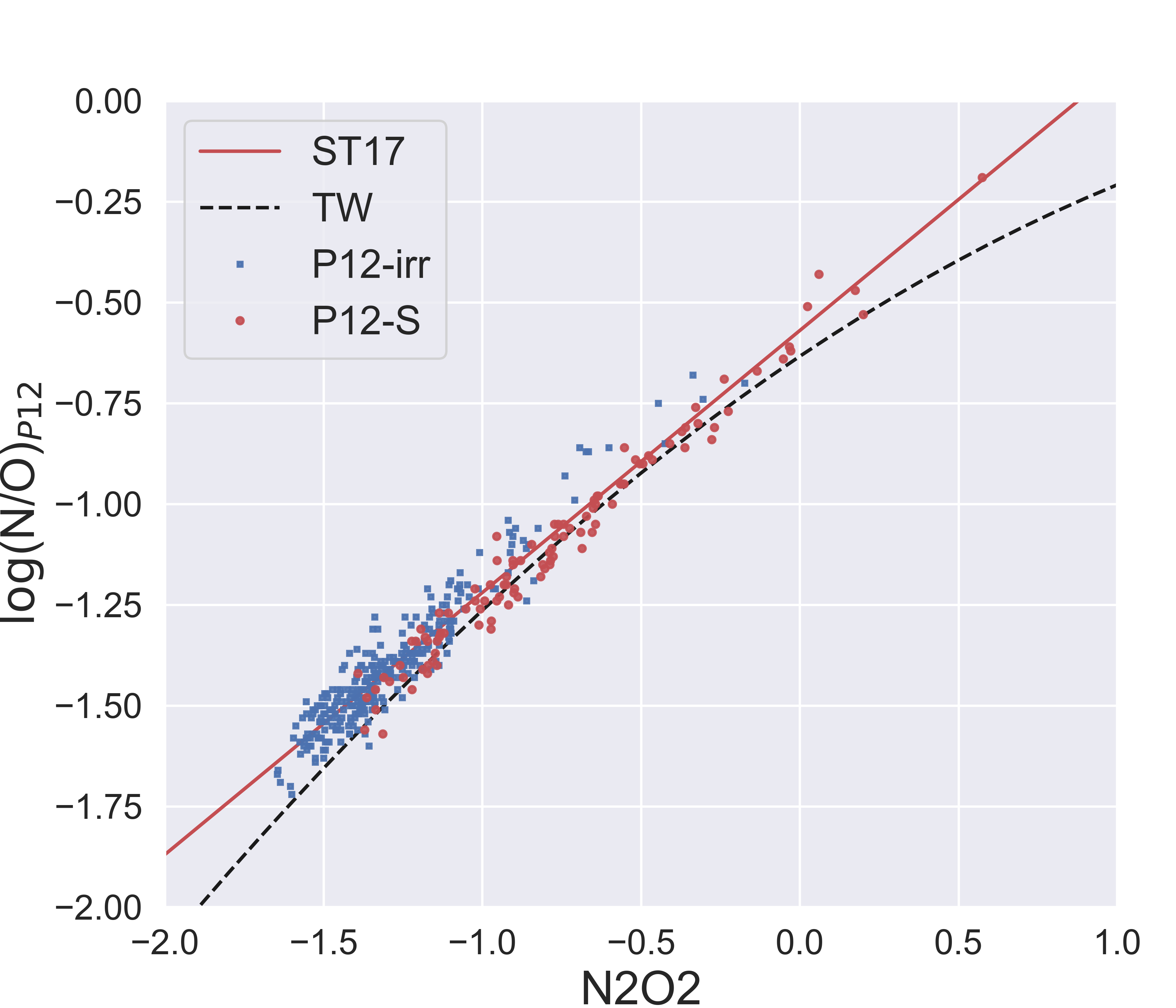}
	\caption{$T_e$-values for  $\log$(N/O) vs. N2O2 index. Abundances are obtained by \citet{Pilyugin2012}. Blue squares are \hii\ regions in irregular galaxies and red dots correspond to \hii\ regions belonging to spiral galaxies.}
	\label{Pilyugin12}
\end{figure}

The \citetalias{Strom2017} calibration is derived from a considerably larger sample than in \citetalias{LS2015}, and contains 414 extragalactic \hii\ regions collected by \citet{Pilyugin2012}, of which only 105 belong to spiral galaxies\footnote{All the \hii\ regions belonging to spiral galaxies in the \citet{Pilyugin2012} sample are included in the \citet{AZ2021a} \hii\ region sample used in this paper.}  and the rest are from irregular galaxies. As we can see in Fig.~\ref{fig-ajustesNO-otros}, the \citetalias{Strom2017} calibration almost matches the one performed in this work (Eq.~\ref{fit1}) in the range -0.8 $\lesssim$ N2O2 $\lesssim$ -0.3 dex, finding the largest difference between ours and \citetalias{Strom2017}'s calibration at the highest and lowest values of N2O2. However, this difference is small, $\lesssim$0.15-0.20 dex. The comparison between \citetalias{Strom2017}'s and our empirical calibration is better seen in Fig.~\ref{Pilyugin12}, where the \citet{Pilyugin2012} \hii\ region sample employed by \citetalias{Strom2017}  is also plotted, with blue squares and red dots for \hii\ regions from irregular and spiral galaxies, respectively. We can see that there are less than 10 \hii\ regions in their calibrating sample with N2O2 greater than -0.3. The differences between their and our calibration might then be produced by their larger concentration of regions in the lowest abundance area (bottom-left part of the plot), with respect to our sample, in which the data are more uniformly distributed. In addition, it is important to note that the \citet{Pilyugin2012} sample is dominated by \hii\ regions from irregular galaxies whose location in the $\log$(N/O)-N2O2 diagram is slightly shifted towards higher N/O for a given N2O2 value, with respect to the average trend for the \hii\ regions in spirals. Our more uniform distribution of values across the N2O2 axis, joined to the absence of \hii\ regions from irregulars in our sample, might be the reason why we find it necessary a second order polynomial to properly fit the $\log$(N/O)-N2O2 relation. We will further discuss this point in Section~\ref{discussion}.

In addition to the empirical calibration of N2O2 based exclusively on \hii\ regions, in Table~\ref{tab:otros} and Fig.~\ref{fig-ajustesNO-otros} we compare our calibration with those by \citet{PMC09}, \citet{LA2020}, \citet{PM2021}, and \citet{Hayden2021}, hereinafter \citetalias{PMC09}, \citetalias{LA2020}, \citetalias{PM2021}, and \citetalias{Hayden2021}, respectively, based on other types of calibrating objects. \citetalias{PMC09} use a sample of 271 \hii\ galaxies, 161 giant extragalactic \hii\ regions and 43 \hii\ regions of the Milky Way and Magellanic Clouds; 475 objects in total. 
The \citetalias{PMC09} calibration clearly deviates from ours, being the offset greater than $\sim$0.5 dex for the highest values of N2O2. The scatter in the $\log$(N/O)-N2O2 plot of the calibration sample in \citetalias{PMC09} is considerably high (see their fig.~10). This, together with their fitting method (a least-squares bisector linear fit) could explain, at least in part, the observed deviation from our best-fitting. 

The calibrations of the N2O2 index performed by \citetalias{LA2020}  and \citetalias{PM2021}  are also based on $T_e$-based N/O abundances, but on these cases for samples of nearby galaxies, selected to be local analogues of higher redshift galaxies. Therefore, their calibrations are intended to be useful for abundance estimates for more distant galaxies. The calibrating sample of \citetalias{LA2020} is composed by 27 galaxies with $z < 0.3$, but with properties similar to Lyman Break Galaxies (LBGs, i.e. luminous, compact and having similar rest-frame far-UV properties to typical LBGs). The one used by \citetalias{PM2021} is much larger, with 1240 EELGs (Extreme Emission Line Galaxies), at redshifts $0.00 < z < 0.49$, extracted from the SDSS-DR7  \citep{SA2010}. Both the \citetalias{LA2020} and \citetalias{PM2021}  calibrations are derived from data within a similar N2O2 range (approximately from $-$1.7 to $-$0.5, cf. from $-$1.7 to $+$0.6 in our calibration for only \hii\ regions). 
However, the \citetalias{LA2020} and \citetalias{PM2021} analytical calibrations of N2O2 have different slopes (Fig.~\ref{fig-ajustesNO-otros}), and this translates in an  increasing difference in the $\log$(N/O) values derived from them as the N2O2 index increases. Thus, $\log$(N/O) as derived from \citetalias{PM2021} is $\sim$0.2 dex larger than the value derived from the \citetalias{LA2020} calibration at the top end of their validity  range (N2O2 $\sim -0.5$). The  \citetalias{LA2020} calibration is in rather good agreement  with our derived ones (Eqs.~\ref{fit1} and ~\ref{fit2}), in spite of the different nature of the calibrating targets. This is clearly visible in Fig.~\ref{fig-ajustesNO-otros}, but also in columns 7 and 8 in Table~\ref{tab:otros} that shows that this calibration, when applied  to our \hii\ region sample, yields some of the smallest values for the median residuals (difference between the $T_e$-based $\log$(N/O) and the $\log$(N/O) values derived from this calibration), 0.09 dex, and for the  standard deviation (0.02 dex; cf. 0.14 and 0.12 dex for median residuals and standard deviation, respectively, for the \citetalias{PM2021} calibration)

\citetalias{Hayden2021} has recently derived an empirical calibration of N2O2 from integrated galaxy spectra. It is based on  $T_e$-based N/O abundances for stacked SDSS spectra of star-forming galaxies at redshift $0.07<z<0.25$  \citep{Curti2017}.  It covers a smaller N2O2 range than ours, between $-1.25$ and $0.1$, and underestimates $\log$(N/O)  for most of the \hii\ regions in our sample, as seen in the median residuals value, which is $-0.06$~dex, although with low standard deviation  (0.09).

The  \citetalias{Strom2018} calibration of N2O2 is also derived from integrated galaxy spectra, but relies on N/O abundances derived from photoionization models \citep[Cloudy,][]{Ferland2013}. The calibration sample comprises 148 high redshift galaxies ($<z> =2.3$) from the Keck Baryonic Structure Survey (KBSS) with index values in the range  $-1.7$ < N2O2 < $-0.2$ dex. This calibration has its larger departure from ours at low values of N2O2, $-1.7$ < N2O2 < $-1.2$, but traces well the $T_e$-based $\log$(N/O)-N2O2 relation for our sample of \hii\ regions for N2O2 $\gtrsim -0.9$.

Finally, we would like to point out that, although not shown here for simplicity, we have carefully analysed the differences between the derived  $\log$(N/O) values for the N2O2 calibrations mentioned above, and the $T_e$-based estimates, in plots similar to the one shown in Fig.~\ref{fig-ajustesNO-N2O2-res}. These differences or residuals show a  clear correlation with the 12+$\log$(O/H) from the $T_e$-based  method for all calibrations analysed in this section.

\subsection{Methods based on the N2S2 and N2 indices}
There are not many calibrations of these indices for the derivation of $\log$(N/O), but the ones available are shown in   Fig.~\ref{NO-N2-N2S2-O3N2}. For the N2S2 index, \citetalias{PMC09}, \citetalias{Strom2018}, \citetalias{PM2021} and \citetalias{Hayden2021} provide calibrations obtained from the samples and methods already described in subsection~\ref{methods-N2O2}.  \citet[][hereinafter   \citetalias{Amorin2010}]{Amorin2010} also derive an empirical calibration of the N2S2 index from $T_e$-based abundances for a sample of star-forming galaxies selected from the SDSS. As for the calibration of N2O2 derived from galaxy integrated spectra, the N2S2 index range covered by these calibrating samples is smaller than the one covered by our \hii\ region sample. The calibrations by \citetalias{Strom2018} and \citetalias{PM2021} are the ones that better match the $T_e$-based abundances of our sample of \hii\ regions, having similar slopes in the $\log$(N/O)-N2S2 plane to ours, while the ones by \citetalias{PMC09}, \citetalias{Amorin2010} and \citetalias{Hayden2021} have a steeper slope. The $\log$(N/O) values obtained from the \citetalias{PMC09} and  \citetalias{Amorin2010} calibrations are, on average, larger than the $T_e$-based ones by $\sim$ 0.24$-$0.31 dex, in the upper bound of their corresponding validity range. This offset is larger than the data standard deviation ($\sim$0.19 dex). The \citetalias{Hayden2021} calibration is valid in a short range of N2S2 values, $-0.5<$N2S2$< 0.3$ (c.f. $-0.58<$N2S2$< 0.84$, for our calibrating \hii\ region sample). Its larger slope implies a slight  overestimation of the $\log$(N/O) predicted values from this method, that is comparable to the standard deviation of the residuals.

\citetalias{Strom2018}, in addition to N2O2 and N2S2, also provide an empirical calibration for N2 that matches very nicely our derived calibration for this index (Fig.~\ref{NO-N2-N2S2-O3N2}) in the validity range in common for the two calibrations, in spite that it was obtained from high redshift galaxies.
To our knowledge there is no published empirical calibration of the O3N2 index for the derivation of $\log$(N/O).

\begin{table*}
\hspace{-1cm}
  \begin{minipage}{\textwidth}
\centering
\caption{List of strong-line calibrations proposed in the literature to derive $\log$(N/O), together with those derived in this work (TW). Columns: (1) Strong-line index or relevant strong-line ratios for that method. (2) Secondary parameter on which  $\log$(N/O) depends for that calibration. (3) Number and (4) nature of the objects used to derive the empirical calibration. (5) Range of validity for the calibration as stated by the different authors. (6) Reference. (7) Standard deviation and (8) median values for the N/O residuals (i.e. difference between the SL method for our sample of \hii\ regions, and those derived from the $T_e$-based method, $\log$(N/O)$_{SL}$ - $\log$(N/O)$_{T_e}$) and (9) method used for the determination of $\log$(N/O) for the calibrating targets in column (4): from the $T_e$-based method ($T_e$), photoionization models (PhM), R and C-methods (R,C) and the \hii-CHI-mistry (HCM) method.
\label{tab:otros}}
{\footnotesize
\begin{tabular}{ccrcccccc} 
\hline 
     Index or  & Secondary  & N       & Calibration$^a$  & Validity$^b$  &   Ref.$^c$    & Residuals   & Median      & Method \\
  strong-line  & parameter  &         & objects     & range     &                    & stdev       & residuals  &        \\
    ratio(s)   &            &         &              & (dex)     &                    & (dex)       & (dex)      &         \\
     (1)       & (2)        & (3)     & (4)          & (5)       &       (6)          &   (7)       & (8)        & (9)     \\
\hline
        N2O2   &             & 536    & \hii\ regions                   &  -1.74 < N2O2 < 0.62  & TW & 0.08 & 0.002  &  $T_e$ \\
        N2O2   & $\log$(O/H)& 536     & \hii\ regions                   &  -1.74 < N2O2 < 0.62  & TW   & 0.04 & -0.004  &  $T_e$ \\
        N2O2   &            & 475     &  \hii\ regions (MW, MC) (43),  &  -2.0 < N2O2 < 0.3   & PM09 & 0.13 & 0.21    & $T_e$ \\
               &            &         & GEHRs (161), \hii\ galaxies (271)&    &  & && \\
        N2O2   &            &  48     & \hii\ regions                   &  -1.45 < N2O2 < 0.15    & LS15& 0.09& 0.04 & $T_e^d$\\
        N2O2   &            & 414     & \hii\ regions (P12)             &  -1.7 < N2O2 < 0.2      & ST17 & 0.09&0.05  & $T_e$ \\
        N2O2   &            & 148     & KBSS galaxies, <$z$>=2.3        &  -1.7 < N2O2 < -0.2     & ST18 & 0.11 & 0.05 & PhM\\
        N2O2   &            &  27     & LBAs ($z$ < 0.3)            &  -1.5 < N2O2 < -0.5     & LA20 & 0.09& 0.02 & $T_e$\\
        N2O2   &            & 1240    & EELGs (<$z$>=0.08)           &  -1.7 < N2O2 < -0.5     & PM21 & 0.12& 0.14 & $T_e$\\
        N2O2   &        & 118478 & SFGs {\em (stacks}, <$z$>=0.072)  & -1.25 < N2O2 < 0.1      & HP21 & 0.09 & -0.06 & $T_e$\\\hline
        N2S2   &            & 530     & \hii\ regions                   & -0.58 < N2S2 < 0.84     & TW     & 0.19 &-0.004 & $T_e$ \\
        N2S2   &            & 475     & \hii\ regions (MW, MC) (43) & -1.4 < N2S2 < 1         & PM09   & 0.21 & 0.24 & $T_e$ \\
               &            &         & GEHRs (161), \hii\ galaxies (271) &    &  &  &  & \\
         N2S2  &            & 148     &  KSBB galaxies, <$z$>=2.3       & -0.6 < N2S2 < 0.3           & ST18   &  0.19 & 0.07 & PhM \\  
         N2S2  &            & --      & SFGs (0.03$\le z \le$ 0.37)  & -0.58 $\le$ N2S2 $\le$-0.02 & A10 & 0.34 & 0.37 & $T_e$\\
         N2S2  &            & 1240    & EELGs (<$z$>=0.08)          &  -0.8 < N2S2 < 0.3          & PM21 & 0.19& 0.06  & $T_e$\\
         N2S2  &            & 118478 & SFGs   {\em (stacks}, <$z$>=0.072)                          &  -0.5 < N2S2 < 0.3 & HP21 & 0.20 & 0.16 & $T_e$ \\\hline
         O3N2  &            &  530    &  \hii\ regions                 &  -1.61 < O3N2 < 2.86            & TW     & 0.15 & 0.03 &  $T_e$\\
         O3N2&$\log$(O$_{32}$)& 530    &  \hii\ regions                 &  -1.61 < O3N2 < 2.86            & TW     & 0.09 & 0.03 &  $T_e$\\\hline
         N2    &             & 294     &  \hii\ regions                &  -1.5 < N2 < -0.65              & TW      & 0.13 & 0.01 & $T_e$\\ 
         N2 &$\log$(O$_{32}$) & 294     & \hii\ regions                 &   -1.5 < N2 < -0.65             & TW     & 0.09 &-0.003 & $T_e$\\
         N2            &     & 148     & KSBB galaxies, <$z$>=2.3      &   -1.6 < N2 < -0.3              & ST18   & 0.15 & -0.02 & PhM\\\hline
         N$_2$, R$_2$  &     & 313     & \hii\ regions                 &  $-0.7\le\log(\rm{N2})\le-0.45$ &PG16&0.09 & 0.03& $T_e$ (R,C)\\   
         N2O2, R$_{23}$ &    & 38 478  & SFGs (0.04$\le z \le$ 0.25) &  $-1.8\le\log$(N/O)$\le-0.1$    & LI06 & 0.09 &  -0.03 &   PhM    \\  
         \rm{[}\oii], [\oiii]& & 550  & \hii\ regions (MW, MC)   &  --                              & PM14  & 0.13 & 0.06 & PhM (HCM) \\
         \rm{[}\nii], [\sii] & &      &  GEHRs, \hii\ galaxies         &                                  &       &      &      &  \\\hline
	\end{tabular}}
{\footnotesize
\begin{tabular}{p{\textwidth}}
$^a$ Acronyms as follow: EELGs (Extreme Emission-Line Galaxies), GEHRs (Giant  Extragalactic \hii\ Regions), KBSS (Keck Baryonic Structure Survey), LBA (Lyman Break analogs), MC (Magellanic Cloud), MW (Milky Way), and SFGs (Star Forming Galaxies). \\ 
$^b$ Validity ranges are approximate and are deduced from the figures in the original papers, except in \citet{PM2021}, \citet{Amorin2010} and \citet{PG2016}, where they are given explicitly by the authors. \\
$^c$ TW: This work; PM09:  \citet{PMC09}, ST17:  \citet{Strom2017}, ST18: \citet{Strom2018},  LA20: \citet{LA2020},  LS15: \citet{LS2015},  A10:  \citet{Amorin2010} ,  LI06:  \citet{Li2006}, PM14:  \citet{PM14},  PG16:  \citet{PG2016}, PM21: \citet{PM2021}, HP21: \citet{Hayden2021}.\\
$^d$ The N/O abundances in \citetalias{LS2015} were derived from an empirical determination of $T_e$, from their derived 12+$\log$(O/H) abundances.
\end{tabular}}
\end{minipage}
\end{table*}


\subsection{Other methods}
There are other methods that permit to estimate $\log$(N/O) from several strong-line ratios. This is the case for the R-calibration and the \hii-CHI-mistry (HCM) method by \citet{PG2016} and \citet{PM14}, respectively.

\citet{PG2016} performed an empirical calibration (the R calibration) after updating the sample of \citet{Pilyugin2012}. It uses the strong-line ratios  $R_2 = I_{[OII]}(\lambda 3727 + \lambda 3729)/I_{H\beta}$ and $N_2 = I_{[NII]}(\lambda 6548 + \lambda 6584)/I_{H\beta}$.
Their calibrating sample is a selection of 313 \hii\ regions (out of 965), chosen because of having O/H and N/O abundances, as derived from a strong line method \citep[C method,][]{Pilyugin2012}, that differ less than 0.1~dex from the $T_e$-based ones.
The calibrating \hii\ region sample belongs to nearby spiral and irregular galaxies, and yields a relation valid in the range $-0.7 \lesssim \log{N_2} \lesssim -0.45$, which corresponds approximately to the same N2O2 interval of the calibration derived in this paper. If we apply the R calibration  to our \hii\ region sample, we obtain a standard deviation of 0.09 dex and a median value of 0.03 dex  for the differences between the R-calibration $\log$(N/O) values  and those derived from the  $T_e$-based method, which are very similar to what we get for empirical calibrations of N2O2 using only \hii\ regions (see Table~\ref{tab:otros}).

The \hii-CHI-mistry method \citep[HCM][]{PM14}, based on a bayesian-like comparison between a certain set of emission-lines and the predictions from a large grid of photoionization models was applied to the same \hii\ sample used in this paper. The results are widely discussed in \citet{AZ2021a}, and also shown in Table~\ref{tab:otros}\footnote{It is important to note that in \citet{AZ2021a}, HCM version 3 was used, that uses POPSTAR synthesis evolutionary models \citep{Molla09}. The HCM results may change for different assumed model, e.g. for BPASS v2.1 models \citep{Eldridge2017}, more appropriated for EELGs as discussed in \citetalias{PM2021}.}. 

Also based on a combination of photoionization models \citep[from][]{Thurston1996} and strong-line ratios ([\nii]$\lambda$6548,6583/[\oii]$\lambda$3727 and R23), \citet{Li2006} performed a calibration based on 38478 star forming galaxies from SDSS. Median value and standard deviation of residuals are very similar to those found with the \citet{PG2016} method.


\section{Discussion}
\label{discussion}

We have derived new empirical calibrations for the estimation of the N/O abundance ratio in local Galactic and extragalactic \hii\ regions, that make use of the strong-line ratios N2O2, N2S2, N2 and O3N2, being the calibration of N2O2 the one that best reproduces the $T_e$-based estimate (Eqs.~\ref{fit1} and \ref{fit2}).
     
\begin{figure}
 \centering
\includegraphics[width=0.5\textwidth]{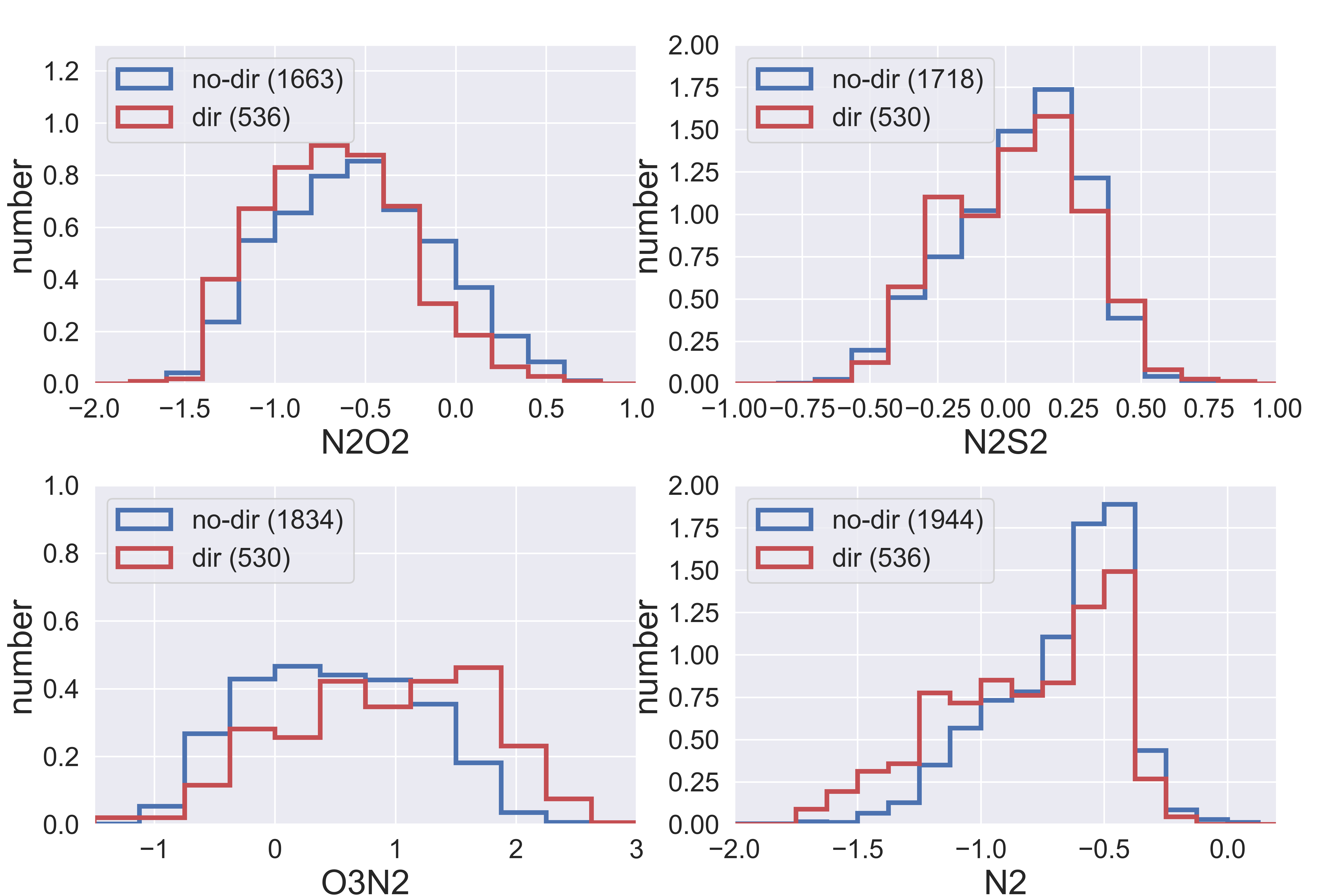}
\caption{Normalized histograms showing the distribution of the indices used in this work (from left to right and from top to bottom N2O2, N2S2, O3N2, N2)  for the \hii\ region calibration sample (in red) and the sample of \hii\ regions without $T_e$-based N/O determinations (blue), both from \citet{AZ2021a}}
\label{hist}
\end{figure}

A key issue for any {\em empirical} strong-line method is the potential effect of sample bias. This occurs when the calibration sample from which $T_e$-based abundances are derived is not representative of the entire sample (including targets without auroral line detections) for which strong-line abundances will be estimated. \citet{Kewley2019} propose a simple exercise to test for sample bias, which is to compare the distribution of relevant emission-line ratios of the calibration sample (with $T_e$-based abundances) with that of the  sample without $T_e$-based abundances.
We show this comparison in Fig.~\ref{hist} and  Table~\ref{tab:distributions} for the \hii\ region collection used in this work \citep{AZ2021a}. Both the \hii\ region calibration sample  (in red) and the sample of regions without $T_e$-based N/O determinations (in blue) cover the same index ranges. The distributions for the two subsamples are very similar for all indices in terms of median values and standard deviations, specially for N2S2  and N2O2. The largest difference between the calibration sample and the sample of regions without $T_e$-based N/O occurs for O3N2, with the median value shifted $\sim$0.5~dex towards larger values of O3N2, but this shift is not very relevant taking into account the width of the corresponding distributions ($\sim$0.8 dex).

\begin{table}
\centering
\caption{Median index value and standard deviation (in dex) for the distributions of  N2O2, N2S2, O3N2 and N2 for the sample of calibration \hii\ regions (columns 2 and 3) and for the sample of \hii\ regions without  $T_e$-based  N/O (columns 4 and  5).}
	\label{tab:distributions}
	\begin{tabular}{lccccccccl}  
\hline
\multicolumn{1}{c}{}   & \multicolumn{3}{c}{Calibration sample}  & \multicolumn{3}{c}{Regions without {\em direct} N/O}\\
\hline
 Index &  Median value & stdev  & & &Median value & stdev \\
       &  (dex)        & (dex)  & & &(dex)        & (dex) \\
    (1) &  (2)         & (3)    & & &(4)          & (5)   \\      
                \hline
                  N2O2    & -0.67  & 0.40  & & &-0.55  & 0.44 \\  
                  N2S2    &  0.06  & 0.25  & & &0.08   & 0.25 \\  
                  O3N2    & 0.97   & 0.82  & & &0.46   & 0.71 \\  
                  N2      & -0.76  & 0.35  & & &-0.61  & 0.27 \\  
		\hline
	\end{tabular}
\end{table}

The effect of sample bias is  well illustrated by comparing our calibration sample with the one used by \citetalias{Strom2017}. As already mentioned in Section~\ref{methods-N2O2}, the N2O2 calibration of \citetalias{Strom2017}  is the only one  that exclusively uses extragalactic \hii\ regions with $T_e$-based N/O abundances as calibrators (apart from ours and that of \citetalias{LS2015} which is based on regions from a single galaxy). They use a  large sample of \hii\ regions from \citet[][\citetalias{Pilyugin2012} hereinafter]{Pilyugin2012} in which most of them (309 out of 414) belong to irregular galaxies, rather than spirals as in our calibration.

\begin{figure}
 \centering
\includegraphics[width=0.5\textwidth]{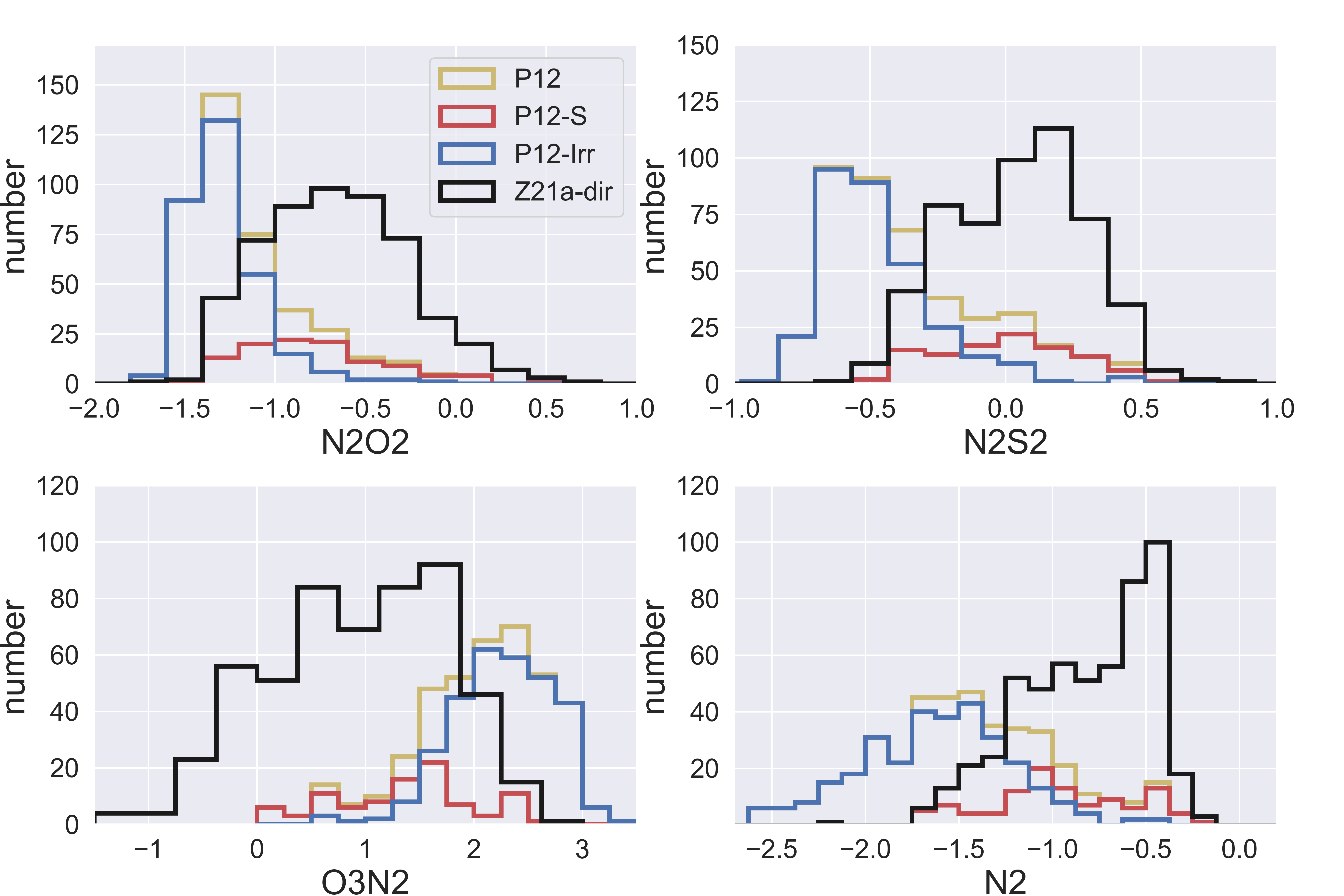}
\caption{Distribution of the indices used in this work to estimate N/O values (from left to right and from top to bottom N2O2, N2S2, O3N2, N2), considering several samples: \hii\ regions with $T_e$-based N/O in \citet{AZ2021a} (Z21a-dir), \hii\ regions with $T_e$-based abundances compiled by \citet{Pilyugin2012} (P12). For the latter we also show the subsamples of \hii\ regions belonging to spiral (P12-S) and irregular (P12-Irr) galaxies.}
\label{hist-P12}
\end{figure}

In Fig.~\ref{hist-P12} we show the distributions of the N2O2, N2S2, O3N2 and N2  indices, for the \citetalias{Pilyugin2012} sample and, separately, for the subsamples of regions  belonging  to  irregular (P12-Irr, in blue) or spiral (P12-S, in red) galaxies in \citetalias{Pilyugin2012}. We also show, for comparison, the distribution for our \hii\ region calibration sample (Z21a-dir). It can be seen that the distributions for \citetalias{Pilyugin2012} (in yellow) and Z21a-dir (in black) are  quite different, despite both being based on \hii\ regions. However, $\sim$75 per cent of the \citetalias{Pilyugin2012} \hii\ region sample  belong to irregular galaxies (P12-Irr), and these regions seem to be producing the differences in the index distributions between the \citetalias{Pilyugin2012} and Z21a-dir samples, as can be seen in the P12-Irr and P12-S comparison. But more important than the differences in the distributions themselves, is the fact that the two samples cover different ranges of index values, being especially evident for the O3N2 index, which covers much lower index values in Z21a-dir.

\begin{figure*}
\centering
\includegraphics[width=1\textwidth]{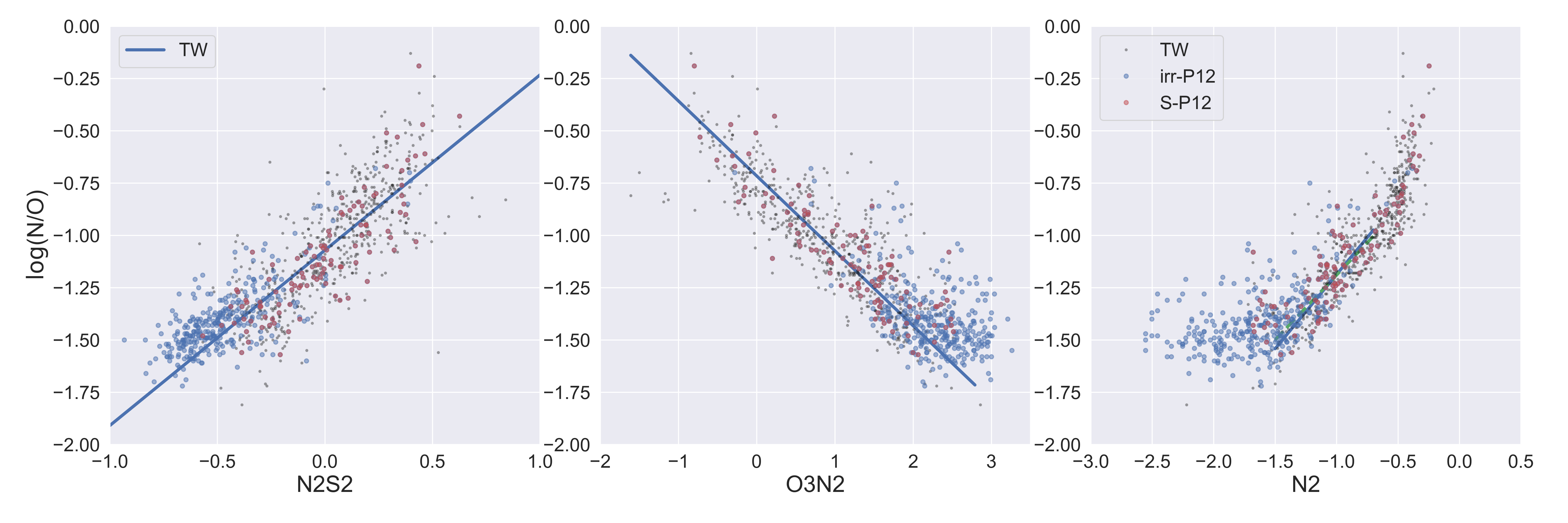}
\caption{The same as Fig.\ref{NO-N2-N2S2-O3N2} but showing the sample of \hii\ regions with $T_e$-based abundances from \citet{Pilyugin2012} with blue or red dots depending on whether they are located in irregular or spiral galaxies, respectively.}
\label{NO-N2-N2S2_O3N2_Py}
\end{figure*}

The differences in the distribution of N2O2 index values might then be the origin of the observed differences at high and low N/O values (Section~\ref{methods-N2O2}) between the N2O2 calibration derived in this work and the one derived by \citetalias{Strom2017}. This is easily seen in Fig. ~\ref{Pilyugin12}, where the $T_e$-based N/O for the  \citetalias{Strom2017} calibration sample (P12) is plotted as a function of N2O2. The larger content of \hii\ regions from irregulars in the \citetalias{Pilyugin2012} sample (blue squares in Fig. ~\ref{Pilyugin12}) and their concentration towards lower values of N2O2, yields a linear $\log$(N/O)-N2O2 relation that differs from the second order polynomial necessary to properly fit the data when the sample is exclusively based on \hii\ regions in spirals (Eq.\ref{fit1}), with a more uniform sampling across the  $\log$(N/O)-N2O2 sequence. 
Therefore, the \citetalias{Strom2017} calibration, if applied to \hii\ regions in spirals in their validity range ($-1.7<$N2O2$<0.2$) would bias the derived strong-line N/O towards larger values for regions with low $\log$(N/O) ($\lesssim -1.25$~dex) or N2O2 ($\lesssim -1.1$). The subsamples P12-Irr and P12-S are also plotted in the  $\log$(N/O) vs. N2S2, O3N2 and N2 diagrams in Fig.~\ref{NO-N2-N2S2_O3N2_Py}. The \hii\ regions from irregulars also produce a bias in them, but the higher dispersion in these relations dilutes their effect.

The differences observed between our and other previous strong-line calibrations of N2O2 to derive N/O, are more difficult to interpret when the calibrating samples are different from \hii\ regions and/or their abundances are obtained from photonization models rather than $T_e$-based determinations. In the lowest N/O range, the calibration of N2O2 with the largest difference with ours is the one derived by \citetalias{Strom2018}, whose calibrators are  high redshift galaxies (<$z$> =2.3) with N/O abundances derived from photoionization models. However, their derived $\log$(N/O)-N2O2 relation shows a reasonably good overlapping with our sample of local \hii\ regions in the range $-1 < $N2O2 $< -0.5$, where \citetalias{Strom2018} have the bulk of their data and these have the lowest scatter  (see their Fig.11). The calibration derived by \citetalias{LA2020} for nearby analogs of Lyman Break Galaxies is also in good agreement with ours in their range of validity. The same is not true for the one obtained by \citetalias{PM2021}  for EELGs, which has a steeper slope than ours and those of \citetalias{Strom2018} and \citetalias{LA2020}. There is a issue that may be relevant to explain the differences between this work and those cited above. Although the redshift range of the \citetalias{PM2021} (0<$z$<0.49) sample overlaps with that of  \citetalias{LA2020} ($z$<0.3), the \citetalias{PM2021} sample  has an average redshift of 0.08 with only 10\% of the galaxies in their sample having $z$>0.194. The 3\arcsec of the SDSS fibre includes flux from the central region of the sample galaxies in \citetalias{PM2021}  ($\sim$5~kpc for their average $z$). The abundances of  \citetalias{Strom2018} and \citetalias{LA2020} come, however, from integrated fluxes from the whole disc. More difficult is the comparison with  \citetalias{Hayden2021}.
Like \citetalias{PM2021}  they consider low redshit galaxies ($z<$0.25, <$z$>=0.072) and use SDSS spectra, with the fibre including only the emission from the central part of the galaxies, as in \citetalias{PM2021}, but in this case the emission-line fluxes has been estimated for stacked SDSS spectra.

The $\log$(N/O)-N2S2 relation has a larger scatter than the one for the N2O2 index, but it is very useful, given the proximity in the spectra of the involved emission lines. There are five calibrations of this index from previous work (shown in Table~\ref{tab:otros} and Fig.~\ref{NO-N2-N2S2-O3N2}). As with the N2O2 calibration, the one obtained by \citetalias{Strom2018}, despite being obtained with integrated spectra of galaxies at  <$z$>=2.3, is in good agreement with ours, while those of \citetalias{Amorin2010}, \citetalias{Hayden2021}  and \citetalias{PMC09}, obtained from more nearby galaxies (as well as \hii\ regions, in the case of \citetalias{PMC09}), show a steeper slope. These similarities in the calibration obtained from \hii\ regions of local galaxies and high-$z$ galaxies was already found by \citetalias{Strom2018}, who pointed out to similarities between the nebular spectra of \hii\ regions and integrated spectra of high-z galaxies as the cause, perhaps because the latter are dominated by one or several dominant \hii\ regions, as opposed to closer galaxies, where star formation is more distributed in the discs \citep[e.g.][]{Sanders2016,Kashino2017,LA2020}. However, there are well known differences between the spectral properties of \hii\ regions and those of high-redshift galaxies (or their local analogs), namely their different location in the [\oiii]$\lambda5007$/\hb\ versus [\nii]$\lambda6584$/\ha\ Baldwin-Phillips-Terlevich \citep[BPT,][]{BPT1981} diagram \citep[e.g.][]{Steidel2014,Strom2017,Strom2018,Steidel2016,Sanders2021}. 

As commented before regarding the differences with the  \citetalias{PM2021} for N2O2, calibrations obtained with nearby galaxies are also usually based on SDSS spectra, which do not include the integrated emission of the whole galaxy, but of the central area, which, depending on its extent and on the galaxy properties, may be more or less representative of the integrated spectrum of the galaxy. The different covering fraction of the galaxy light in integrated spectra of galaxies may cause part of the observed differences between different calibrations  \citep[e.g.][]{Kewley2005,Li2006}. However, this argument does not explain the reasonably good agreement of our calibration with the one from \citetalias{PM2021} for N2S2, which includes only a part of the light emitted by the galaxies, given the mean redshift (<$z$>=0.08) of their sample.
  
With respect to integrated spectra, an additional aspect, which may be of great relevance to the abundances of the ionised gas, among others, is the possible inclusion of emission from the diffuse ionised gas (DIG)  \citep[e.g.][]{Zurita2000,Oey2007} in spectra that include emission from all or part of the galaxy \citep[see e.g.][]{Sanders2017,Zhang2017,Vale-Asari2019}, as the emission-line ratios in this low surface brightness component differ from those measured in classical \hii\ regions. The DIG could therefore differentialy affect the indices used to trace abundances when these are derived from integrated spectra that includes both \hii\ region and DIG emission.  The [\sii]/\ha\ emission is enhanced in the DIG (more than that of [\nii]/\ha\ ratio\footnote{In fact, the N2O2 is the least sensitive strong-line diagnostic to the DIG and AGN emission \citep{kewley06,Zhang2017}.}) with respect to the \hii\ regions \citep[e.g.][]{Galarza99,Domgorgen1994}. Therefore, in the integrated spectra of galaxies, we would expect a lower N2S2 for a given N2O2 than for individual \hii\ regions.

\begin{figure}
\centering
\includegraphics[width=0.45\textwidth]{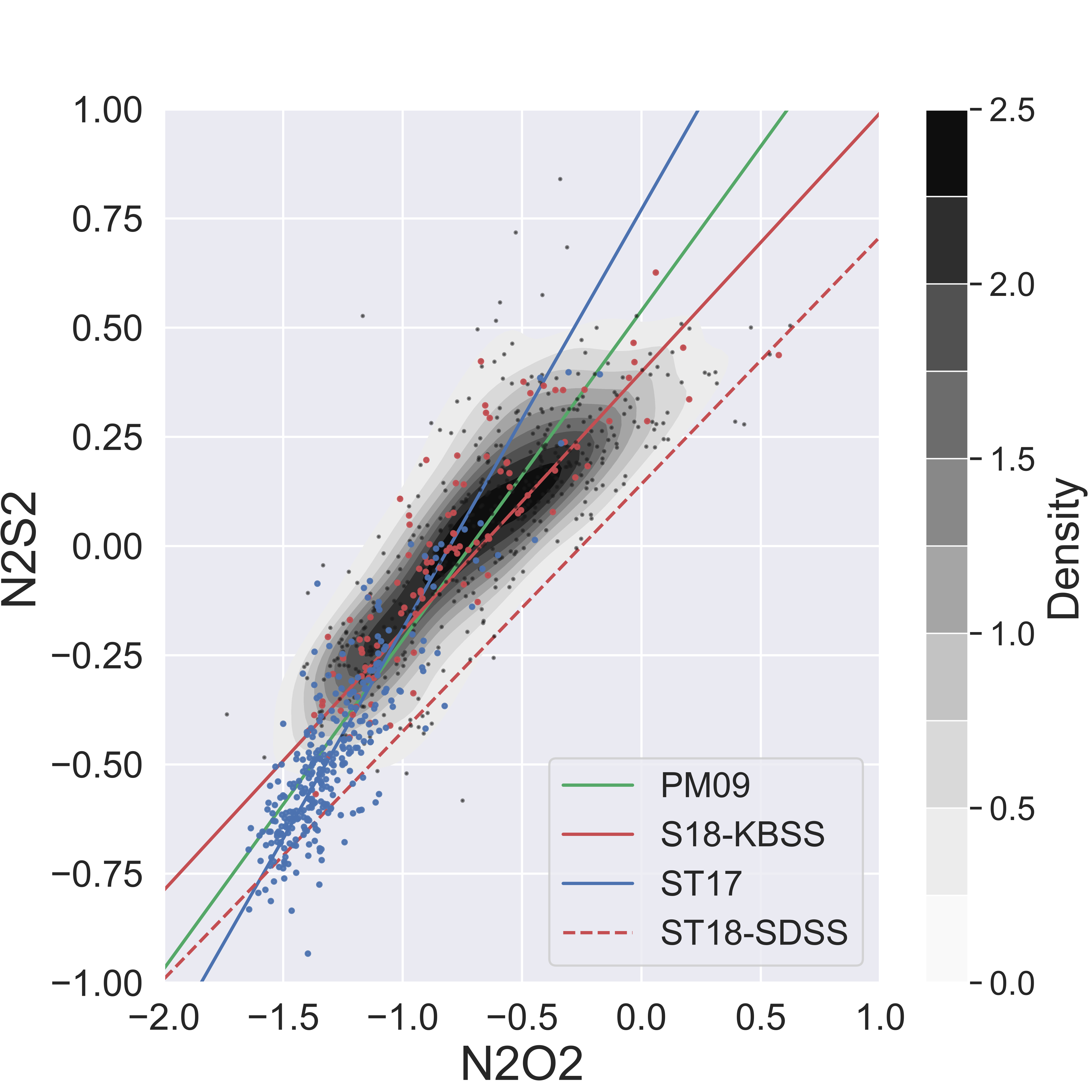}
\caption{N2S2 versus N2O2 for the calibration sample from the  \hii\ region calibration sample used in this work \citep{AZ2021a}, represented as a grey-scale density map and with black dots. We also show the best linear fits derived by other authors for different data samples: those of  \citet{Strom2018} for KBSS galaxies at <z>$\sim$2, and for local SDSS galaxies (red solid and dashed lines, respectively); that of \citet{PMC09} for \hii\ galaxies and \hii\ regions (green line), and the relation derived by \citet{Strom2017}  for the \hii\ regions sample of  \citet{Pilyugin2012} (blue solid line). The \hii\ regions of the \citet{Pilyugin2012} are also shown with blue or red dots depending on whether they belong to irregular or spirals galaxies, respectively.}
\label{N2S2-N2O2}
\end{figure}

Fig.~\ref{N2S2-N2O2} shows (as a density map) the ratio of the N2S2 and N2O2 indices for our sample of \hii\ regions in local spirals \citep{AZ2021a}  and, for comparison, the relations between these two indices derived by previous authors: namely, the relation derived by \citetalias{Strom2018} for high-redshift galaxies (ST18-KBSS), those derived by the same authors for nearby galaxies from SDSS spectra (ST18-SDSS), and for the \citetalias{Pilyugin2012} \hii\ region sample (in local spirals and irregulars) and the relation derived by \citetalias{PMC09}, including \hii\ galaxies and \hii\ regions. It is clearly seen that for a given N2O2, the N2S2 relation derived from the integrated spectrum of local galaxies (ST18-SDSS) is lower  by $\sim$0.2-0.25~dex. This is expected in the case of contamination of DIG emission  in the spectra. However, this does not seem to affect the integrated spectra of high-redshift galaxies. A possible explanation is a decreasing contribution of the DIG component with increasing redshift. As discussed by \citet{Sanders2017}, high-redshift galaxies have both a smaller size and a higher SFR, at a fixed M$_\star$, than their low-redshift analogs. Therefore, they have both a higher specific SFR and \ha\ surface brightness. The mean DIG emission fraction  decreases with increasing \ha\ surface brightness of galaxies \citep{Oey2007}.  The agreement in the N2S2-N2O2 ratio for resolved \hii\ regions and integrated outflows of high-redshift galaxies could be due to a much smaller impact of the DIG on the integrated fluxes of these galaxies.
Finally, we observe again in Fig.~\ref{N2S2-N2O2} the different location and trend of the \hii\ regions of irregular galaxies, which explains the steeper slope of the N2S2-N2O2 relation obtained by \citetalias{Strom2017}. 

There are a number of additional methods to derive N/O, that make use of a combination of the same or other ratios of strong emission lines (bottom rows in Table~\ref{tab:otros}), namely the ones by  \citet{Li2006}, \citet{PM14}, and that from \citet{PG2016}.  We have applied these methods to our \hii\ region calibration sample and they give a good overall prediction of the $T_e$-based N/O abundances, with low values for the median and standard deviation for the differences between the strong-line and the $T_e$-based abundances for all the regions. However, it should be noted that the results obtained with the analytical calibration of the N2O2 index derived in this work give similar or even better results than these methods, with the added advantage of being straightforward to apply.

\begin{figure}
\centering 
 \includegraphics[width=0.47\textwidth]{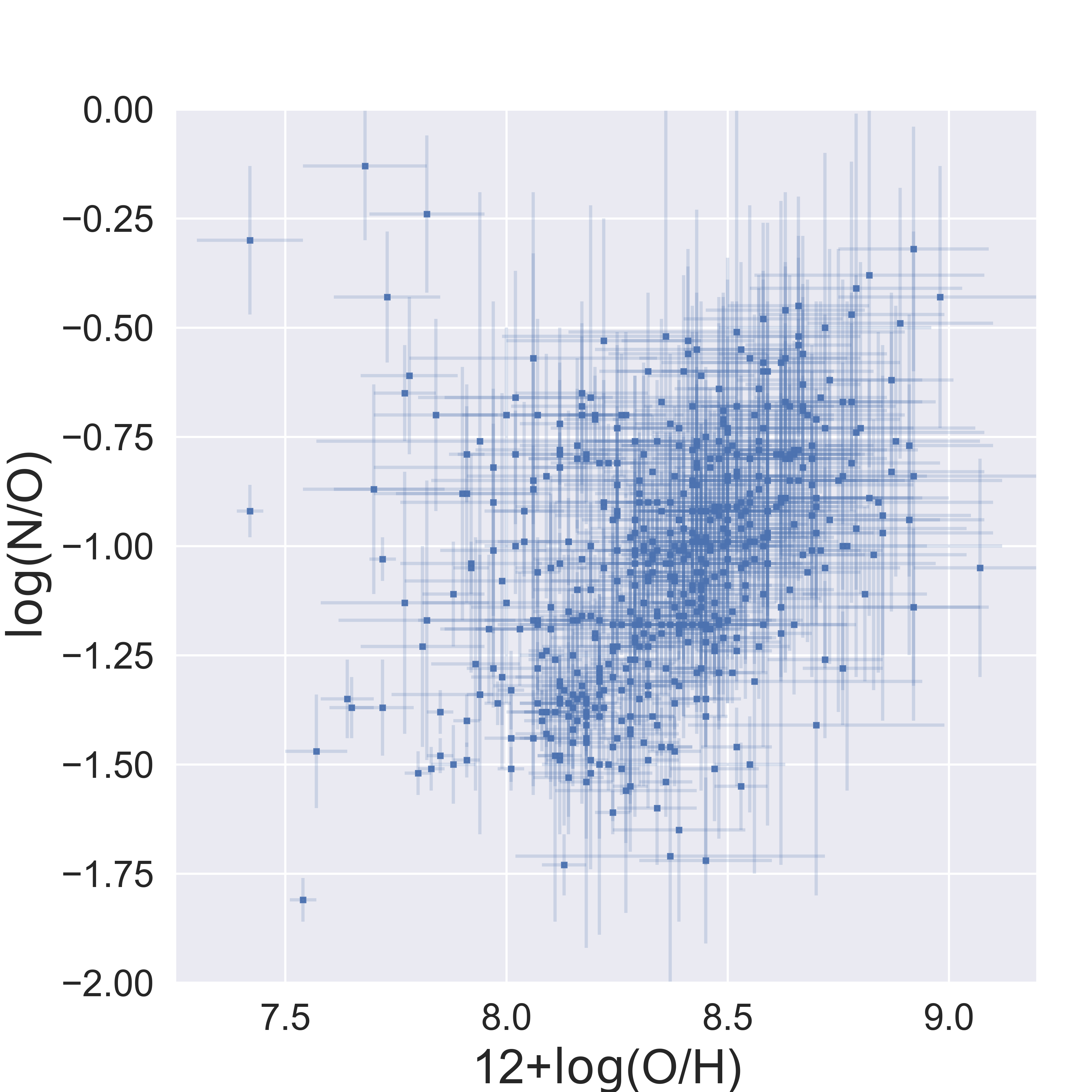}
 \caption{$\log$(N/O) vs. $12+\log$(O/H) for our \hii\ region calibration sample or regions with $T_e$-based N/O and O/H abundances in \citet{AZ2021a}. The median error values are: 0.17 dex for $\log$(N/O) and 0.1 dex for 12+$\log$(O/H).}
\label{NO-OH}
\end{figure}

A final issue that it is worth mentioning here regards the N/O-O/H relation. Some calibrations rely in a assumed relation between $\log$(N/O)-$\log$(O/H) \citep[see e.g.][]{Dopita2016}. The indices used in this paper as tracers of the N/O abundance ratio in \hii\ regions, are also used as metallicity tracers (see Section~\ref{calib} and references therein). A tight relation in the $\log$(N/O)-$\log$(O/H) relation would easily explain their usefulness for the two abundance ratios. However the  $\log$(N/O)-$\log$(O/H), in Fig.~\ref{NO-OH}, for our sample of \hii\ regions with $T_e$-based abundances, has a large dispersion $\gtrsim 0.26$~dex, three times larger than the dispersion in the  $\log$(N/O)-N2O2 relation for the same sample of \hii\ regions (see Table~\ref{tab:ajustes}). This fact strengthens the goodness of our N2O2 calibration, which is not determined by a tight N/O-O/H relation.
 
\section{Conclusions}
\label{conclusions}
We have used a large sample of 536 \hii\ regions in neaby spiral galaxies with $T_e$-based ({\em direct}) abundances to test the usefulness of some of the most frequently used strong-line ratios that involve the [\nii]$\lambda6583$ emission-line flux as tracers of the  N/O abundance ratio in \hii\ regions.
The calibration \hii\ region sample  is extracted from the compilation performed by \citet{AZ2021a}, for which  $T_e$-based abundances were calculated from the compiled emission-line fluxes with a homogeneous methodology. The four indices analysed (N2O2, N2S2, O3N2 and N2) are strongly correlated with the $T_e$-based $\log$(N/O), with Spearman's rank correlation coefficients larger than 0.8 (c.f. $\sim0.6$ with  $12+\log$(O/H)). The strongest correlation is found for the N2O2 index ($\rho$=0.95), with the best fitting function for the log N/O-N2O2 relation being a second-order polynomial:

\begin{equation}
\label{fit_con}
 \begin{array}{rl}
     \log\mbox{(N/O)}_{N2O2} = & (-0.102 \pm 0.018) \times \rm{N2O2}^2    \\
                        & +(0.528 \pm 0.019) \times \rm{N2O2} - (0.634 \pm 0.006) \\
 \end{array}
\end{equation}
This relation is valid in the range $-1.74 <$ N2O2 $< 0.62$, which implies a wide range of $\sim$1.1 dex in $\log$(N/O), with low dispersion in the fitting residuals ($rms \sim0.09$). These show a positive correlation with the $T_e$-based oxygen abundance that, if taken into account in the fitting (Eq.~\ref{fit2} and Table~\ref{tab:ajustes}), reduces the dispersion down to $\sim0.04$~dex.

The dependence of the $T_e$-based $\log$(N/O) on the other three indices (N2S2, O3N2 and N2) fits well with a single linear fit, but the dispersion of the residuals is larger ($rms \sim 0.13 - 0.20$~dex) than for N2O2, and their range of validity smaller, especially for N2 that saturates for high values of N2 ($\gtrsim -0.65$). These results, joined to its virtually independence on the ionization parameter, make N2O2 the preferred index for deriving $\log$(N/O) in \hii\ regions when the electron temperature is not available.

Although these relations have been derived for \hii\ regions in local galaxies, there is good agreement between our calibrations and the ones derived for integrated fluxes of high redshift galaxies by previous authors \citepalias[e.g.][in the range of validity in common]{Strom2018, LA2020}. The N2O2 index is also one of the indices less altered by contamination from the diffuse ionized gas emission \citep[e.g.][]{Zhang2017} and it is virtually independent on the ionization parameter \citep[e.g.][]{Dopita2000}. These properties of the N2O2 index, joined to the observed agreement between our calibration of the N2O2 and previous calibrations of this index for the integrated emission of high-redshift galaxies in the common validity range, make the N2O2 calibration derived in this paper tentatively promising for N/O estimates for integrated fluxes of high-redshift galaxies. Caution is however mandatory until proper calibrations based on sufficient $T_e$-based N/O abundances for high-redhift galaxies (or analogs) become available.

\section*{Acknowledgements}

We thank the  anonymous referee for his/her constructive report, that improved the clarity of the manuscript. This work is supported by the Spanish 'Ministerio de Ciencia e Innovaci\'on' and from the European Regional Development Fund (FEDER) via grants PID2020-224414GB-I00 and PID2020-113689GB-I00, and from the 'Junta de Andaluc\'{\i}a' (Spain) local government through the FQM108 and A-FQM-510-UGR20 projects. EPM acknowledges financial support from the State Agency for Research of the Spanish MCIU through the 'Center of Excellence Severo Ochoa' award to the Instituto de Astrof\'isica de Andaluc\'ia (SEV-2017-0709) and to projects 'Estallidos7' PID2019-107408GB-C44 (Spanish 'Ministerio de Ciencia e Innovaci\'on'), and from the 'Junta de Andaluc\'ia' Excellence project EXC/2011 FQM-7058, and  also  the assistance from his guide dog Rocko without whose daily help this work would have been much more difficult.

\section*{Data Availability}

The data underlying this article are available in the article and are based on a compilation performed by Zurita et al. 2021a that is publicly available on the CDS VizieR facility (via https://vizier.u-strasbg.fr/viz-bin/VizieR).




\bibliographystyle{mnras}
\bibliography{Florido} 




\bsp	
\label{lastpage}
\end{document}